\documentclass[prd,preprint,showpacs,preprintnumbers,amsmath,amssymb,eqsecnum]{revtex4}

\usepackage{bm}

\newcommand{\bv}{\bar{\varphi}}
\newcommand{\conn}{\bar{\Gamma}^{k}_{ij}}
\newcommand{\chris}{{\Gamma}^{k}_{ij}}
\newcommand{\bF}{\bar{F}}
\newcommand{\intp}[1]{\int\frac{d^n{#1}}{(2\pi)^n}}

\begin{document}

\title{Quantum gravity, gauge coupling constants, and the cosmological constant}

\author{David J. Toms}
\homepage{http://www.staff.ncl.ac.uk/d.j.toms}
\email{d.j.toms@newcastle.ac.uk}
\affiliation{%
School of Mathematics and Statistics,
Newcastle University,
Newcastle upon Tyne, United Kingdom, NE1 7RU}

\date{\today}

\begin{abstract}
The quantization of Einstein-Maxwell theory with a cosmological constant is considered. We obtain all logarithmically divergent terms in the one-loop effective action that involve only the background electromagnetic field. This includes Lee-Wick type terms, as well as those responsible for the renormalization group behaviour of the electric charge (or fine structure constant). Of particular interest is the possible gauge condition dependence of the results, and we study this in some detail. We show that the traditional background-field method, that is equivalent to a more traditional Feynman diagram calculation, does result in gauge condition dependent results in general. One resolution of this is to use the Vilkovisky-DeWitt effective action method, and this is presented here. Quantum gravity is shown to lead to a contribution to the running charge not present when the cosmological constant vanishes. This re-opens the possibility, suggested by Robinson and Wilczek, of altering the scaling behaviour of gauge theories at high energies although our result differs. We show the possibility of an ultraviolet fixed point that is linked directly to the cosmological constant.
\end{abstract}


\maketitle

\section{Introduction}

Einstein gravity when quantized about a fixed background (for example flat space) is not renormalizable~\cite{tHooftVeltman,DeservanN1,DeservanN2,DeservanN3,DeserTsaovanN}.  The basic reason for this is that the gravitational coupling constant $G$ has units of inverse mass squared in natural ($\hbar=c=1$) units.  From the standard quantum field theory point of view this means that when working to higher orders in perturbation theory the degree of divergence of diagrams must increase with the order that one is working to.  Naively, we expect a behaviour like $(G\Lambda_c^2)$ to some positive power, where $\Lambda_c$ is a momentum cutoff with the power increasing with the number of loops.

The natural energy scale is set by the Planck mass $M_P=(\hbar c/G)^{1/2}\sim 10^{19}$GeV.  Provided that we restrict ourselves to energies $E\ll M_P$ it is expected that an effective field theory treatment of Einstein's theory is valid.  Indeed classical general relativity is well tested, so we know that quantum effects must be very small.  The methodology for realizing this is the effective field theory framework.  Its application to gravity was emphasized by Donoghue~\cite{Donoghue1,Donoghue2}. (See \cite{Burgess} for a comprehensive and readable review.) What is it expected is that any fundamental theory should give the same results as quantization of Einstein's theory plus matter fields at energies below the Planck scale.  We will concentrate on quantization of Einstein-Maxwell theory is an example.

Robinson and Wilczek~\cite{RobWilczek} presented a calculation that claimed quantum gravity could alter the behaviour of running gauge coupling constants in Yang-Mills theory. Their calculation showed that the renormalization group $\beta$-function receives a purely quantum gravitational contribution that tends to render all theories asymptotically free, irrespective of what happens in the absence of gravity. The phenomenological consequences of their calculation were examined in \cite{Gogoladze}, and in addition attracted attention from possible applications~\cite{Huang} to the weak gravity effect~\cite{ArkaniHamed,Banks}. In view of the potential importance of the Robinson-Wilczek result a number of independent examinations were undertaken.

Doubt was first cast on the Robinson-Wilczek conclusion by Pietrykowski~\cite{Piet} who showed that their result was gauge condition dependent. By choosing a different gauge no quantum gravitational correction to the $\beta$-function was found. Because of the question of gauge condition dependence, a subject that will be studied in depth later in the present paper, we undertook a gauge condition independent calculation~\cite{DJT1} and supported the conclusion of Pietrykowski; in pure Einstein-Maxwell theory the $\beta$-function receives no contribution from quantum gravity. Dimensional regularization~\cite{tHooft3} was used in \cite{DJT1}, and this is only sensitive to logarithmic divergences. Because the quantum gravity calculation of \cite{RobWilczek} involved quadratic divergences, the role of regularization dependence of the result was studied~\cite{Ebertetal} in Einstein-Yang-Mills theory. By using both a momentum space cut-off, and ensuring gauge invariance by applying the Taylor-Slavnov-Ward-Takahashi identities~\cite{Taylor,Slavnov,Ward,Takahashi}, it was shown~\cite{Ebertetal} that the quadratic divergences cancelled and that the result agreed with what was found using dimensional regularization. No purely quantum gravitational contribution to the $\beta$-function was found in agreement with \cite{Piet,DJT1}. A further analysis \cite{TangWu} showed that it was possible to find a regularization scheme that could result in a non-zero gravitational contribution to the $\beta$-function, although the relation with previous work mentioned is unclear at this point. More recent work has examined the applications to Yukawa and $\phi^4$ interactions~\cite{RodSchustnew} (see also \cite{Perc}) and to higher dimensions~\cite{EbertetalJHEP}. Implications for the Lee-Wick~\cite{LeeWick1,LeeWick2} mechanism for gravity have also been considered~\cite{Wu1,Wu2,RodSchust}. It is also worth noting that a string calculation~\cite{Kiritsis} in a supersymmetric model results in no gravitational correction to the $\beta$-function.

In contrast to the negative results found for pure gravity, we showed~\cite{DJT2} that if a cosmological constant was present, then a non-zero quantum gravitational correction to the $\beta$-function could be obtained, that was different from what Robinson and Wilczek~\cite{RobWilczek} found, but that still tended to result in asymptotic freedom. One purpose of the present paper is to give more details of the calculation described in \cite{DJT2}. Another is to extend the calculation to the poles in the effective action that involve higher derivatives of the electromagnetic field, including those of the Lee-Wick type. A third is to show that when calculated using traditional background-field methods, or equivalently using standard Feynman rules, the pole terms calculated do depend on the choice of gauge condition. This will be illustrated by explicit calculation below. The gauge condition independent background-field method due to Vilkovisky~\cite{Vilkovisky1,Vilkovisky2} and DeWitt~\cite{DeWittVD} will be used, and dimensional regularization adopted. This method is outlined in Sec.~\ref{VD} and applied to Einstein-Maxwell theory in the subsequent sections. We can make a brief comment on quadratic divergences at this stage to justify the use of dimensional regularization. It is possible to show that the quadratic divergences are completely independent of the Vilkovisky-DeWitt correction to the traditional background-field formalism. Thus the quadratic divergences will agree with what is found using a traditional Feynman diagram calculation and cancel as found in \cite{Ebertetal}. Only logarithmic divergences will survive and these are calculable by dimensional regularization.

\section{The gauge independent effective action\label{VD}}
\subsection{Introduction\label{VD1}}

In the quantization of any gauge theory there are two main problems to be addressed.  The first is that the results must be invariant under the underlying gauge transformations that define the theory.  Within the background-field method this is relatively easy to do~\cite{DeWittDynamical,Honerkamp,tHooftgauge,DeWittQGII,Boulware}.  A classic paper showing how this works in Yang-Mills theory is Abbott's~\cite{Abbott} calculation of the $\beta$-function to two-loop order.  Within a more traditional Feynman diagram calculation, gauge independence is guaranteed by the Slavnov-Taylor-Ward-Takahashi identities satisfied by the various $n$-point functions~\cite{Ward,Takahashi,Taylor,Slavnov}.  It is therefore possible to ensure gauge invariance of the calculation, even after regularization.

The second problem that must be overcome concerns the possible dependence of the results on the choice of gauge condition.  Within the context of the functional integral approach to the background-field method, the gauge condition must be introduced to avoid over-counting field configurations that are related by gauge transformations in the integration over the space of all fields.  This is usually dealt with by the imposition of a gauge condition and the associated ghost fields, the Faddeev-Popov~\cite{FaddeevPopov} method. The choice of gauge condition is arbitrary, and it is at this stage that the dependence on this arbitrary choice can enter the calculation.  If we focus on the computation of the effective action using the background-field method, then the effective action can become dependent upon the choice of gauge condition.

An early example that illustrates the dependence of the effective action on the gauge condition is the calculation of the effective potential (a special case of the effective action) in scalar quantum electrodynamics at one-loop order by Dolan and Jackiw~\cite{DolanJackiw1}.  The one-loop effective potential was shown to depend explicitly on parameters used to implement the gauge condition.  A later computation by Dolan and Jackiw~\cite{DolanJackiw2} showed that the one-loop effective potential computed in the unitary gauge differed from that previously calculated.  This gauge condition dependence can affect physically measurable quantities, such as the critical temperature in finite-temperature field theories, so is not a problem that can be ignored.  Often the gauge condition independence is obscured in calculations because a convenient choice of gauge condition is made to expedite the calculations, and all trace of the parameters disappears.  This does not solve the problem, merely hides it.

A key feature of the background-field method that leads to a possible dependence on the gauge conditions at one-loop order is that it is necessary to expand the field about an arbitrary background field that is not the solution to the classical equations of motion.  (After all, one motivation for the use of the effective potential in gauge theories was to study symmetry breaking due to radiative corrections by minimizing the effective potential to determine the ground state~\cite{ColemanWeinberg}.  This is not the same as the effective potential evaluated at a classical solution.) It is possible to modify the background-field method as discussed by Nielsen~\cite{Nielsen} for scalar electrodynamics to obtain a result for the effective potential that does not depend on the choice of gauge condition, thereby ensuring that physical consequences of the theory do not depend on this choice.  However, another approach is more direct: modify the background-field method at the start to ensure that the effective action is independent of gauge condition.  This modification was suggested originally by Vilkovisky~\cite{Vilkovisky1,Vilkovisky2} and refined by DeWitt~\cite{DeWittVD} to apply to all orders in the loop expansion, and it is this approach that we will adopt here.  A brief outline of some of the more important features for the calculations needed in this paper follow in the next section.  (A more pedagogical review can be found in \cite{ParkerTomsbook}.)

\subsection{Vilkovisky-DeWitt effective action\label{VDeffectiveaction}}

The use of DeWitt's condensed index notation~\cite{DeWittDynamical} is almost indispensable here.  We will consider only bosonic gauge fields denoted by the generic symbol $\varphi^i$.  Here $i$ stands for all of the normal gauge indices, spacetime indices, as well as the dependence on the spacetime coordinates.  Repeated indices are summed over in the usual way in the Einstein summation convention, but in addition carry an integration over the included spacetime coordinates.  Let $S\lbrack\varphi\rbrack$ represent the classical action functional for the theory.  We assume that the theory has a gauge invariance that can be described using infinitesimal parameters $\delta\epsilon^\alpha$.  (Again $\alpha$ is a condensed index.) We will assume that the infinitesimal gauge transformation can be written as
\begin{equation}\label{VD2.1}
\delta\varphi^i=K^{i}_{\alpha}\lbrack\varphi\rbrack\delta\epsilon^\alpha
\end{equation}
for some functional $K^{i}_{\alpha}\lbrack\varphi\rbrack$ that can be regarded as the generator of gauge transformations.  (We will be more explicit about what $K^{i}_{\alpha}\lbrack\varphi\rbrack$ is in the next subsection.) Invariance of the action functional $S\lbrack\varphi\rbrack$, that is $S\lbrack\varphi+\delta\varphi\rbrack=S\lbrack\varphi\rbrack$ holds to first order in $\delta\epsilon^\alpha$, results in
\begin{equation}\label{VD2.2}
K^{i}_{\alpha}\lbrack\varphi\rbrack S_{,i}\lbrack\varphi\rbrack=0
\end{equation}
where $S_{,i}\lbrack\varphi\rbrack$ denotes the functional derivative of $S\lbrack\varphi\rbrack$ with respect to $\varphi^i$.  Hamilton's principle of stationary action tells us that $S_{,i}=0$ are the classical equations of motion; thus, (\ref{VD2.2}) expresses the fact that these equations are invariant under a gauge transformation.

We have already mentioned the problem of quantization of gauge theories using the integration method over the space of all fields (the Feynman path integral).  If we naively integrate over the space of all gauge fields we will include fields as different even though they are physically equivalent under the gauge transformation (\ref{VD2.1}).  We can think of all fields related by gauge transformations as belonging to the same equivalence class and we wish to integrate in the functional integral only over distinct equivalence classes.  The first step in the implementation of this is to introduce a gauge condition (sometimes call the gauge-fixing condition)
\begin{equation}\label{VD2.3}
\chi^\alpha\lbrack\varphi\rbrack=0.
\end{equation}
We require $\chi^\alpha\lbrack\varphi+\delta\varphi\rbrack=\chi^\alpha\lbrack\varphi\rbrack$ hold only if $\delta\epsilon^\alpha=0$. The consequence of this is that
\begin{equation}\label{VD2.4}
Q^{\alpha}{}_{\beta}\lbrack\varphi\rbrack\delta\epsilon^\beta=0
\end{equation}
has only the solution $\delta\epsilon^\beta=0$ where we have defined
\begin{equation}\label{VD2.5}
Q^{\alpha}{}_{\beta}\lbrack\varphi\rbrack=\chi^{\alpha}{}_{,i}\lbrack\varphi\rbrack K^{i}_{\beta}\lbrack\varphi\rbrack\;.
\end{equation}
Provided that $\det Q^{\alpha}{}_{\beta}\ne0$, (\ref{VD2.4}) does imply that $\delta\epsilon^\beta=0$ is the only solution as required. ($\det Q^{\alpha}{}_{\beta}$ is the Faddeev-Popov~\cite{FaddeevPopov} factor that we will return to later.) Note also that the gauge condition can depend on the background field, although we will not indicate this dependence explicitly.

The next step in the Vilkovisky-DeWitt effective action relies on assuming that the space of all fields is equipped with a metric tensor $g_{ij}\lbrack\varphi\rbrack$.  We can write a line element as usual.  In the case of Yang-Mills theory and gravity there are natural choices that do not involve the introduction of dimensional parameters as we will discuss below.  (For gravity, the result is the DeWitt metric~\cite{DeWittmetric}.) For both gravity and Yang-Mills theory it is possible to show that $K^{i}_{\alpha}\lbrack\varphi\rbrack$ can be viewed as components of a set of vector fields that form a Lie algebra and are Killing vectors for the field space metric $g_{ij}$.  (See \cite{ParkerTomsbook} for details.)

The central part of the Vilkovisky approach is the choice of connection.  One way to calculate the appropriate connection is by first considering a general displacement in the space of fields $d\varphi^i$.  This will not be generated by a gauge transformation in general, but will be expressible as a linear combination
\begin{equation}\label{VD2.6}
d\varphi^i=\omega_{\parallel}^{i}+\omega_{\perp}^{i},
\end{equation}
where
\begin{equation}\label{VD2.7}
\omega_{\parallel}^{i}=K^{i}_{\alpha} d\epsilon^\alpha,
\end{equation}
and $\omega_{\perp}^{i}$ satisfies
\begin{equation}\label{VD2.8}
g_{ij}\omega_{\perp}^{i}\omega_{\parallel}^{j}=0\;.
\end{equation}
To obtain $\omega_{\perp}^{i}$ we can define a projection operator
\begin{equation}\label{VD2.9}
P^{i}{}_{j}=\delta^{i}_{j}-K^{i}_{\alpha}\gamma^{\alpha\beta}K_{\beta j},
\end{equation}
where $K_{\beta j}=g_{ji}K^{i}_{\beta}$ as usual, and $\gamma^{\alpha\beta}$ is the inverse of
\begin{equation}\label{VD2.10}
\gamma_{\alpha\beta}=K^{i}_{\alpha}g_{ij}K^{j}_{\beta}.
\end{equation}
It is easy to verify that
\begin{equation}\label{VD2.11}
P^{i}{}_{j}K^{j}_{\alpha}=0,
\end{equation}
and that
\begin{equation}\label{VD2.12}
P^{i}{}_{j}P^{j}{}_{k}=P^{i}{}_{k}.
\end{equation}
Because of (\ref{VD2.11}), $P^{i}{}_{j} $ has the property of projecting vectors perpendicular to the generators of gauge transformations.  This results in the line element
\begin{eqnarray}
ds^2&=&g_{ij}d\varphi^i d\varphi^j\nonumber\\
&=&g^{\perp}_{ij}\omega_{\perp}^{i}\omega_{\perp}^{j}+\gamma_{\alpha\beta}d\epsilon^\alpha d\epsilon^\beta\label{VD2.13}
\end{eqnarray}
that exhibits the local product structure with the first term on the right hand side representing the line element on the space of orbits and the second term representing that on the gauge group.  In (\ref{VD2.13}) we have
\begin{equation}\label{VD2.14}
g^{\perp}_{ij}=P^{k}{}_{i}P^{l}{}_{j}g_{kl}
\end{equation}
interpreted as the metric on the space of distinct gauge orbits.

Because it is the space of distinct gauge orbits that is integrated over in the Feynman functional integral, the natural choice of connection $\conn$ is determined from the requirement that
\begin{equation}\label{VD2.15}
\bar{\nabla}_i g^{\perp}_{jk}=0=g^{\perp}_{jk,i}-\bar{\Gamma}^{l}_{ij}g^{\perp}_{lk} -\bar{\Gamma}^{l}_{ik}g^{\perp}_{jl}.
\end{equation}
($\bar{\nabla}_i$ denotes the covariant derivative with respect to the connection.)
This leads to
\begin{equation}\label{VD2.16}
\bar{\Gamma}^{l}_{ij}g^{\perp}_{lk}=\frac{1}{2}\left(g^{\perp}_{jk,i}+g^{\perp}_{ki,j} -g^{\perp}_{ij,k}\right).
\end{equation}
Normally we would introduce the inverse to $g^{\perp}_{lk}$ and multiply both sides of (\ref{VD2.16}) by this inverse; however, $g^{\perp}_{lk}$ is not invertible on the full field space since $g^{\perp}_{ij}K^{j}_{\alpha}=0$.  Because of this, $\conn$ is only determined up to an arbitrary multiple of $K^{k}_{\alpha}$ that vanishes when contracted with $g^{\perp}_{lk}$.  It can be shown that $\conn$ takes the form
\begin{equation}\label{VD2.17}
\conn=\chris+T^{k}_{ij}+K^{k}_{\alpha}A^{\alpha}_{ij},
\end{equation}
where $\chris$ is the Christoffel connection for the metric $g_{ij}$, $T^{k}_{ij}$ is a complicated expression that involves $g_{ij},\ K^{i}_{\alpha}$ and their first derivatives, and $A^{\alpha}_{ij}$ is completely arbitrary.

At this stage we note that the effective action can be computed in the loop expansion where in place of the normal derivatives that occur we use covariant ones. (That is, a covariant Taylor expansion of the classical action is used.)  When this is done, it is possible to show that the terms arising from $A^{\alpha}_{ij}$ in (\ref{VD2.17}) vanish as a consequence of gauge invariance; thus, the arbitrariness of the connection is not a problem. Only the Christoffel connection $\chris$ and the term $T^{k}_{ij}$ make a contribution to the result.

When we perform the integration over the space of fields, the natural measure follows formally from (\ref{VD2.13}) as
\begin{equation}\label{VD2.18}
d\mu\lbrack\varphi\rbrack=\left(\prod_{i}\omega_{\perp}^{i}\right) \left(\prod_{\alpha}d\epsilon^\alpha\right)\left(\det g^{\perp}_{ij}\right)^{1/2}\left(\det\gamma_{\alpha\beta}\right)^{1/2}.
\end{equation}
As a consequence of Killing's equation and the anti-symmetric property of the structure constants it is possible to show that $\left(\det g^{\perp}_{ij}\right)^{1/2}$ and $ \left(\det\gamma_{\alpha\beta}\right)^{1/2}$ are both gauge invariant (independent of the gauge parameters $\epsilon^\alpha$).  Thus, if we integrate any gauge invariant expression using the measure (\ref{VD2.18}) the integration over the gauge group parameters $\epsilon^\alpha$ may be factored out leaving only an integration over the orbit space as required. (By orbit space we mean the full field space factored out by the group of gauge transformations.)  Note however that the factor $ \left(\det\gamma_{\alpha\beta}\right)^{1/2}$ remains.  This geometric observation~\cite{EKTFP} is the basis of the usual Faddeev-Popov ``ansatz''~\cite{FaddeevPopov}.  It is now possible show that we can take (with the integration over $\epsilon^\alpha$ dropped but $ \left(\det\gamma_{\alpha\beta}\right)^{1/2}$ kept)
\begin{equation}\label{VD2.19}
d\mu\lbrack\varphi\rbrack=\left(\prod_{i}d\varphi^{i}\right) \left(\det g^{\perp}_{ij}\right)^{1/2} \left(\det Q^{\alpha}{}_{\beta}\right)\delta\lbrack\chi^\alpha\rbrack,
\end{equation}
since the space of orbits is also fixed by the gauge condition $\chi^\alpha=0$.  This corresponds exactly to the usual Faddeev-Popov~\cite{FaddeevPopov} construction.

It is now possible to prove three things about the effective action.  The first is that if we define the standard functional integral expression, expressed in a suitably covariant formulation, it does not depend on the choice of field variables $\varphi^i $ that are chosen.  The second is that the effective action is a gauge invariant functional of the background field.  The third is that the effective action is not dependent on the choice made for the gauge condition.  In proving this last property, the Vilkovisky-DeWitt connection (\ref{VD2.17}) is essential, and in particular the role of $T^{k}_{ij}$ is crucial.  This has been verified explicit calculations~\cite{FradkinTseytlin,BV,HKLT}.

The basic idea now is to pick a gauge choice, compute all the geometric arsenal described above, and calculate the effective action.  Because we are guaranteed that the result does not depend on the choice of gauge condition we can make the calculations simpler by adopting a suitable gauge choice.  This choice was called the Landau-DeWitt gauge by Fradkin and Tseytlin~\cite{FradkinTseytlin} and is sometimes called the background field gauge.  It begins by expressing the field
\begin{equation}\label{VD2.20}
\varphi^i=\bv^i+\eta^i
\end{equation}
where $\bv^i$ is the background field.  The Landau-DeWitt gauge condition reads
\begin{equation}\label{VD2.21}
\chi_{\alpha}=K_{\alpha i}\lbrack\bv\rbrack\eta^i=0.
\end{equation}

Because of the form taken by $T^{k}_{ij}$ it is possible to show that it makes no contribution to the effective action at one-loop order if the Landau-DeWitt gauge is used. Any other choice of gauge requires the inclusion of $T^{k}_{ij}$.  This leads to considerable technical simplifications.  Beyond one-loop order this is no longer the case in general.  For certain classes of theories, including Yang-Mills theory but not gravity, it is possible to prove that $T^{k}_{ij}$ makes no contribution to the effective action to all orders in the loop expansion for the Landau-DeWitt gauge.  Thus the correct gauge invariant and gauge condition independent effective action for Yang-Mills theory can be calculated from the usual formalism provided that we adopt only the Landau-DeWitt gauge; for any other choice of gauge we must use the full Vilkovisky-DeWitt expression~\cite{FradkinTseytlin,Rebhan,ParkerTomsbook}.  We will only use the Landau-DeWitt gauge condition here.

The aim of this paper is to study only quantum corrections to quantum gravity at one-loop order.  This involves an expansion of the classical action in a covariant Taylor series to quadratic order in the quantum field $\eta^i$ defined in (\ref{VD2.20}) followed by a Gaussian functional integral.  The complication due to the presence of the $\delta$-function in the measure (\ref{VD2.19}) can be dealt with by use of the familiar identity
\begin{equation}\label{VD2.22}
\delta\lbrack\chi^\alpha\rbrack=\lim_{\xi\rightarrow0}(4\pi i\xi)^{-1/2}\exp\left(\frac{i}{2\xi}\chi^\alpha\chi_\alpha\right)
\end{equation}
suitably generalized to the case of functions.  The result for the effective action to one-loop order may be taken as
\begin{eqnarray}
\Gamma\lbrack\bv\rbrack&=&S\lbrack\bv\rbrack-\ln\det Q_{\alpha\beta}\lbrack\bv\rbrack\label{VD2.23}\\
&&+\frac{1}{2}\lim_{\xi\rightarrow0}\ln\det\left(\nabla^i\nabla_j S\lbrack\bv\rbrack + \frac{1}{2\xi}K^{i}_{\alpha}\lbrack\bv\rbrack K^{\alpha}_{j}\lbrack\bv\rbrack\right).\nonumber
\end{eqnarray}
We work in the Landau-DeWitt gauge as discussed.  Here,
\begin{equation}\label{VD2.24}
\nabla_i\nabla_j S\lbrack\bv\rbrack=S_{,ij}\lbrack\bv\rbrack - \conn S_{,k}\lbrack\bv\rbrack
\end{equation}
gives the covariant derivative computed using the connection (\ref{VD2.17}).  It should be clear why the arbitrary third term in (\ref{VD2.17}) does not matter at one-loop order (since $K^{k}_{\alpha}S_{,k}=0$ is the expression of gauge invariance).  It is not immediately obvious that the term $T^{k}_{ij}$ makes no contribution in the Landau-DeWitt gauge but it can be shown not to.  (See the pedagogical treatment in \cite{ParkerTomsbook}.) We may therefore replace $\conn$ in (\ref{VD2.24}) with the Christoffel connection $\chris$.

At this stage it should be clear why it is significant to know whether or not we are expanding about a background field that is the solution to the classical equations of motion.  If we are, then $S_{,i}=0 $ and the terms in the effective action that arise from the connection vanish.  The formalism reduces to the usual one.  As we will see, we must not assume that this is the case in what follows.  Another observation that can be made is that if the Christoffel connection vanishes, then by adopting the Landau-DeWitt gauge there is no distinction between covariant and ordinary derivatives, and the usual traditional effective action formalism can be used.  (This occurs in the case where the metric $g_{ij}$ on the space of fields does not depend on the fields.)

\section{Einstein-Maxwell theory}\label{EinMax}

The interest of the present paper is to study the one-loop quantization of Einstein-Maxwell theory as a simple model of a gauge theory coupled to gravity. The classical action functional may be chosen to be
\begin{equation}\label{VD3.1}
S=S_M+S_G,
\end{equation}
where
\begin{equation}\label{VD3.2}
S_M=\frac{1}{4}\int d^nx|g(x)|^{1/2}F_{\mu\nu}F^{\mu\nu},
\end{equation}
is the Maxwell field action, and
\begin{equation}\label{VD3.3}
S_G=-\frac{2}{\kappa^2}\int d^nx |g(x)|^{1/2} (R-2\Lambda),
\end{equation}
is the gravitational Einstein-Hilbert action with the inclusion of a cosmological constant $\Lambda$. We have defined
\begin{equation}\label{VD3.4}
{\kappa^2}=32\pi G,
\end{equation}
with $G$ Newton's gravitational constant, allowed the spacetime dimension to be $n$, and adopted the curvature conventions of \cite{MTW} but with a Riemannian (as opposed to a Lorentzian) metric chosen. There is no deep significance to be attached to this last choice; it merely avoids factors of $i$.

In Sec.~\ref{VD} condensed notation has been used with $\varphi^i$ standing for all of the fields. Although convenient for discussing basic formalism, for practical calculations normal notation must be resorted to. We will make the association
\begin{equation}\label{VD3.5}
\varphi^i=\left(g_{\mu\nu}(x),A_{\mu}(x)\right).
\end{equation}
Here $A_{\mu}$ is the electromagnetic gauge field with the convention
\begin{equation}\label{VD3.6}
F_{\mu\nu}=\partial_{\mu}A_{\nu}-\partial_{\nu}A_{\mu}.
\end{equation}
The Vilkovisky-DeWitt formalism has been set up to be completely covariant. Any choice of field variables ($g^{\mu\nu},A^\mu,|g|^{1/4}g^{\mu\nu}$ etc.) may be made in place of (\ref{VD3.6}) without affecting the results. We have merely adopted the simplest, and perhaps most natural, choice here.

The action (\ref{VD3.1}) is invariant under combined spacetime coordinate changes and $U(1)$ gauge transformations. If we let $\delta\epsilon^\lambda$ be the infinitesimal parameters describing spacetime coordinate transformations, and $\delta\epsilon$ be the infinitesimal parameter for the $U(1)$ gauge transformation, then the fields (\ref{VD3.5}) behave like
\begin{eqnarray}
\delta g_{\mu\nu}&=&-\delta\epsilon^\lambda g_{\mu\nu,\lambda}-\delta\epsilon^{\lambda}{}_{,\mu}g_{\lambda\nu} -\delta\epsilon^{\lambda}{}_{,\nu}g_{\mu\lambda}\;,\label{VD3.7}\\
\delta A_\mu&=&-\delta\epsilon^\nu A_{\mu,\nu}-\delta\epsilon^{\nu}{}_{,\mu}A_\nu +\delta\epsilon_{,\mu}\;.\label{VD3.8}
\end{eqnarray}
These last two results are represented by $\delta\varphi^i=K^{i}_{\alpha}\delta\epsilon^\alpha$ in condensed notation. (See (\ref{VD2.1}).) We will make the condensed index association $\delta\epsilon^\alpha=(\delta\epsilon^\lambda(x),\delta\epsilon(x))$. The indices in (\ref{VD2.1}) can be uncondensed by writing
\begin{eqnarray}
\delta g_{\mu\nu}(x)&=&\int d^nx'\left\lbrace K^{g_{\mu\nu}(x)}{}_{\lambda}(x,x')\delta\epsilon^\lambda(x') +K^{g_{\mu\nu}(x)}(x,x')\delta\epsilon(x')  \right\rbrace,\label{VD3.9}\\
\delta A_{\mu}(x)&=&\int d^nx'\left\lbrace K^{A_{\mu}(x)}{}_{\lambda}(x,x')\delta\epsilon^\lambda(x') +K^{A_{\mu}(x)}(x,x')\delta\epsilon(x')  \right\rbrace.\label{VD3.10}
\end{eqnarray}
Here we use the actual field as a component label as in \cite{Kunstatter87}. By comparing (\ref{VD3.9}) and (\ref{VD3.10}) with (\ref{VD3.7}) and (\ref{VD3.8}) we can read off
\begin{eqnarray}
K^{g_{\mu\nu}(x)}{}_{\lambda}(x,x')&=&-g_{\mu\nu,\lambda}(x)\delta(x,x') -g_{\mu\lambda}(x)\partial_{\nu}\delta(x,x')-g_{\lambda\nu}(x)\partial_{\mu}\delta(x,x'), \label{VD3.11}\\
K^{g_{\mu\nu}(x)}(x,x')&=&0\label{VD3.12}\\
K^{A_{\mu}(x)}{}_{\lambda}(x,x')&=&-A_{\mu,\lambda}(x)\delta(x,x')-A_{\lambda}(x)\partial_{\mu}\delta(x,x'), \label{VD3.13}\\
K^{A_{\mu}(x)}(x,x')&=&\partial_{\mu}\delta(x,x'). \label{VD3.14}
\end{eqnarray}
Here $\delta(x,x')$ is the symmetric Dirac $\delta$-distribution defined by $\int d^nx'\delta(x,x')F(x')=F(x)$ for scalar test function $F(x)$. $\delta(x,x')$ transforms like $|g(x')|^{1/2}$ at $x'$ and a scalar at $x$.

The natural line element for the space of fields is
\begin{equation}
ds^2=\int d^nx d^nx'\left\lbrace g_{g_{\mu\nu}(x)g_{\lambda\sigma}(x')}dg_{\mu\nu}(x)dg_{\lambda\sigma}(x') +g_{A_{\mu}(x)A_{\nu}(x')}dA_{\mu}(x)dA_{\nu}(x')\right\rbrace\label{VD3.15}
\end{equation}
where we choose
\begin{equation}
g_{g_{\mu\nu}(x)g_{\lambda\sigma}(x')}=\frac{1}{2\kappa^2}|g(x)|^{1/2}\left(g^{\mu\lambda}g^{\nu\sigma} + g^{\mu\sigma}g^{\nu\lambda}-g^{\mu\nu}g^{\lambda\sigma}\right)\delta(x,x'),\label{VD3.16}
\end{equation}
to be the DeWitt metric~\cite{DeWittmetric}, and
\begin{equation}\label{VD3.17}
g_{A_{\mu}(x)A_{\nu}(x')}=|g(x)|^{1/2}g^{\mu\nu}(x)\delta(x,x').
\end{equation}
The factor of $\kappa^{-2}$ in (\ref{VD3.16}) ensures that both terms in (\ref{VD3.15}) have the same units and results in $ds^2$ in (\ref{VD3.15}) having units of length squared.

Given the metric components in (\ref{VD3.16}) and (\ref{VD3.17}), the Christoffel connection can be computed. The non-zero components turn out to be
\begin{eqnarray}
\Gamma^{g_{\lambda\tau}(x)}_{g_{\mu\nu}(x')g_{\rho\sigma}(x'')}&=&\vphantom{|}\left\lbrack-\delta^{(\mu}_{(\lambda}g^{\nu)(\rho} \delta^{\sigma)}_{\tau)}+\frac{1}{4}g^{\mu\nu}\delta^{\rho}_{(\lambda}\delta^{\sigma}_{\tau)} \right.\nonumber\\
&&\hspace{-1cm} -\frac{1}{2(2-n)}\left\lbrace g_{\lambda\tau}g^{\mu(\rho}g^{\sigma)\nu} -\frac{1}{2}g_{\lambda\tau}g^{\mu\nu}g^{\rho\sigma}\right\rbrace\nonumber\\
&&\vphantom{|}\left.\hspace{-1cm}+ \frac{1}{4}g^{\rho\sigma}\delta^{\mu}_{(\lambda}\delta^{\nu}_{\tau)} \right\rbrack \delta(x'',x)\delta(x'',x')\label{VD3.18}\\
\Gamma^{g_{\mu\nu}(x)}_{A_{\lambda}(x')A_{\tau}(x'')}&=& \frac{1}{2}\delta^{(\lambda}_{\mu}\delta^{\tau)}_{\nu} \delta(x,x')\delta(x',x''),\label{VD3.19}\\
\Gamma^{A_{\mu}(x)}_{A_{\nu}(x')g_{\alpha\beta}(x'')} &=& \frac{1}{4}\left(\delta^{\nu}_{\mu}g^{\alpha\beta} -\delta^{\alpha}_{\mu}g^{\nu\beta} -\delta^{\beta}_{\mu}g^{\nu\alpha}\right)\delta(x,x')\delta(x,x'')\label{VD3.20}\\
&=&\Gamma^{A_{\mu}(x)}_{g_{\alpha\beta}(x'')A_{\nu}(x')}.\nonumber
\end{eqnarray}
The round brackets around indices denote a symmetrization over the indices enclosed along with a factor of $1/2$. The Christoffel connection components in (\ref{VD3.18}--\ref{VD3.20}) will be used to compute the second term of (\ref{VD2.24}).

At this stage we may choose a background.  If we keep the background metric general then we would be repeating the monumental calculation of \cite{DeservanN2} using the Vilkovisky-DeWitt formalism.  Although this would be interesting and challenging to do, we will focus instead on the quantum gravity corrections to the running value of the electric charge, or fine structure constant, as well as computing the pole terms of the Lee-Wick type.  This means that we do not need to consider terms in the effective action that involve the curvature and we will choose the background spacetime to be flat.  We therefore choose $\bv^i$ to be
\begin{equation}\label{VD3.21}
\bv^i=(\delta_{\mu\nu},\bar{A}_{\lambda}(x)),
\end{equation}
where we keep the background gauge field $\bar{A}_{\lambda}(x)$ general.  If we are only interested in the terms in the effective action that can affect the electric charge then we can take the background electromagnetic field $\bar{F}_{\mu\nu}$ to be constant as in our earlier work~\cite{DJT1,DJT2}; however, this would miss out any poles that involve derivatives of the electromagnetic field that could be of the Lee-Wick type.  We do not make any assumptions about $\bar{A}_{\lambda}(x)$ at this stage.  An important feature of the background is that it is not a solution to the classical Einstein-Maxwell equations, and therefore the inclusion of the connection term in (\ref{VD2.24}) is crucial if the result for the effective action is to be gauge condition independent.  We will illustrate that this is so by an explicit calculation showing how a gauge condition dependent result is obtained using the traditional effective action method.

The results for $S_{,i}\lbrack\bv\rbrack$ can be computed from appropriate functional derivatives of (\ref{VD3.1}--\ref{VD3.3}) with respect to $\varphi^i $ in (\ref{VD3.5}) followed by setting $\varphi^i = \bv^i$ in (\ref{VD3.21}).  The results are
\begin{eqnarray}
\left.\frac{\delta S}{\delta g_{\mu\nu}(x)}\right|_{\bv}&=&\frac{2}{\kappa^2}\Lambda\delta^{\mu\nu} +\frac{1}{8}\delta^{\mu\nu}\bar{F}_{\alpha\beta}\bar{F}^{\alpha\beta} -\frac{1}{2}\bar{F}^{\mu}{}_{\lambda}\bar{F}^{\nu\lambda},\label{VD3.22}\\
\left.\frac{\delta S}{\delta A_{\mu}(x)}\right|_{\bv}&=& \partial_{\nu}\bar{F}^{\mu\nu}.\label{VD3.23}
\end{eqnarray}
The last result (\ref{VD3.23}) vanishes if we restrict $\bar{F}_{\mu\nu}$ to be constant, so would not contribute to the charge renormalization, but can contribute to Lee-Wick type terms.  The second and third terms of (\ref{VD3.22}) are just those involved in the stress-energy-momentum tensor of the electromagnetic field.

It is worth explaining at this stage why we can concentrate on pole terms in the effective action that involve only the electromagnetic field to deduce the charge renormalization.  The basic reason is the Ward-Takahashi identity that relates the charge and field renormalization factors.  The calculation of Abbott~\cite{Abbott} showed how this works within the background field method.  Let $e_B$ and $e_R$ be the bare and renormalized charges respectively.  Using dimensional regularization~\cite{tHooft3,tHooftRG} we have
\begin{equation}\label{VD3.24}
e_B=\ell^{n/2-2}Z_ee_R,
\end{equation}
where $\ell$ is an arbitrary unit of length (the reciprocal of `t~Hooft's~\cite{tHooftRG} unit of mass) and $Z_e$ is the charge renormalization factor.  Similarly, let $\bar{A}_{\mu B}$ and $\bar{A}_{\mu R}$ be the bare and renormalized background gauge fields respectively.  Then,
\begin{equation}\label{VD3.25}
\bar{A}_{\mu B}=\ell^{2-n/2}Z_A^{1/2}\bar{A}_{\mu R},
\end{equation}
with $Z_A$ the field renormalization factor.  A consequence of the gauge invariant background field method is
\begin{equation}\label{VD3.26}
e_B\bar{A}_{\mu B}=e_R\bar{A}_{\mu R}.
\end{equation}
(Think of the gauge covariant derivative written in terms of the bare quantities in the bare classical action.  This must be rewritten in terms of the renormalized ones in a gauge invariant way.) From (\ref{VD3.24}--\ref{VD3.26}) we find
\begin{equation}\label{VD3.27}
Z_eZ_A^{1/2}=1,
\end{equation}
as the Ward-Takahashi identity tells us.  The standard `t~Hooft~\cite{tHooftRG} approach to the renormalization group relates the running value of the charge to the pole terms in $Z_e$. This will be outlined in Sec.~\ref{VD6}.  The identity (\ref{VD3.27}) allows us to deduce the pole terms in $Z_e$ from those in $Z_A$, and $Z_A$ is determined by the renormalization of the background gauge field.  It is for this reason that we can concentrate on the pole parts of the effective action that involve the background gauge field.

The calculation become simpler to deal with if we re-express the last two terms of the one-loop effective action (\ref{VD2.23}) as functional integrals.  The $\ln\det Q_{\alpha\beta}$ term can be expressed as an integration over ghost fields, whereas the last term can be written as an integration over the quantum field $\eta^i$ defined in (\ref{VD2.20}).  We will write
\begin{eqnarray}
\Gamma_G&=&
\frac{1}{2} \ln\det\left\lbrace \nabla^i\nabla_j S\lbrack\bv\rbrack +\frac{1}{2\xi}K^{i}_{\alpha}\lbrack\bv\rbrack K^{\alpha}_{j}\lbrack\bv\rbrack\right\rbrace\nonumber\\
&=&-\ln\int\lbrack d\eta\rbrack\,e^{-S_q}\;,\label{VD3.28}
\end{eqnarray}
where
\begin{equation}\label{VD3.29}
S_q=\frac{1}{2}\eta^i\eta^j\left(S_{,ij}-\Gamma^{k}_{ij}S_{,k} +\frac{1}{2\xi}K_{\alpha\,i}K^{\alpha}_{j}\right),
\end{equation}
with the limit $\xi\rightarrow0$ understood to enforce the Landau-DeWitt gauge condition and all terms evaluated at the background field $\bv$ defined in (\ref{VD3.21}).  The ghost contribution is
\begin{eqnarray}
\Gamma_{GH}&=&-\ln\det Q_{\alpha\beta}\nonumber\\
&=&-\ln\int\lbrack d\bar{\eta}d\eta\rbrack e^{-\bar{\eta}_\alpha Q^{\alpha}{}_{\beta}\eta^\beta},\label{VD3.30}
\end{eqnarray}
where $\bar{\eta}_\alpha$ and $\eta^\beta$ are anti-commuting ghost fields.

The aim now is to identify terms in the arguments of the exponentials of the functional integrals that depend on the background gauge field, treat these terms interactions, and expand in powers of the interactions up to a given order.  Simple power counting shows that there will be poles that involve two, three, or four powers of the background electromagnetic field.  The terms with three powers of the field would be expected to vanish because the classical theory is symmetric under $\bar{F}_{\mu\nu}\rightarrow-\bar{F}_{\mu\nu}$, and we will verify that this is the case below.  We will first concentrate on the gravity and gauge field contribution in $\Gamma_G$ (\ref{VD3.28},\ref{VD3.29}) in the next section.  The ghost contribution will be studied in the subsequent section.

\section {Gravity and gauge field contribution}\label{VD4}
\subsection{Expansion of the effective action}\label{effactgauge}

In evaluating the result for $S_q$ in (\ref{VD3.29}) it can be noted initially that the first term, $\frac{1}{2}\eta^i\eta^jS_{,ij}\lbrack\bv\rbrack$ is just the quadratic term in the Taylor series expansion of $S\lbrack\bv+\eta\rbrack$ .  This is the term (along with that from the gauge condition) that is present in the traditional effective action.  The term that involves the connection is only present in the Vilkovisky-DeWitt approach.  In order to trace the effect of including the connection, we will include a parameter $v$ in $\conn$ that when set to zero gives us the traditional result, and when set to unity gives us the (correct) Vilkovisky-DeWitt result.

To deal with the gauge-fixing condition it can be noted first that the condensed index expression for the gauge-fixing condition
\begin{equation}\label{VD4.1}
S_{GF}=\frac{1}{4\xi}\eta^i\eta^jK_{\alpha i}K^{\alpha}_{j}=\frac{1}{4\xi}(\chi_\alpha)^2
\end{equation}
where $\chi_\alpha$ is the Landau-DeWitt gauge condition (\ref{VD2.21}).  In our case there are two gauge conditions, one for the graviton field, and one for the electromagnetic field.  With the gauge transformation generators given in (\ref{VD3.11}--\ref{VD3.14}) we find the Landau-DeWitt gauge conditions specified by
\begin{eqnarray}
\chi_\lambda&=&\frac{2}{\kappa}(\partial^\mu h_{\mu\lambda}-\frac{1}{2}\partial_\lambda h)+\omega(\bar{A}_\lambda \partial^\mu a_\mu+a^\mu\bar{F}_{\mu\lambda}),\label{VD4.2}\\
\chi&=&-\partial^\mu a_\mu,\label{VD4.3}
\end{eqnarray}
where we have set
\begin{equation}\label{VD4.4}
\eta^i=(\kappa h_{\mu\nu},a_\mu),
\end{equation}
so that
\begin{eqnarray}
g_{\mu\nu}&=&\delta_{\mu\nu}+\kappa h_{\mu\nu},\label{VD4.5}\\
A_\mu&=&\bar{A}_\mu+a_\mu,\label{VD4.6}
\end{eqnarray}
and defined
\begin{equation}\label{VD4.6b}
h=h^{\mu}_{\mu}=\delta^{\mu\nu}h_{\mu\nu}.
\end{equation}
The factor of $\kappa$ in (\ref{VD4.5}) is a standard convenience that removes a factor of $\kappa^{-2}$ present in the Einstein-Hilbert action (\ref{VD3.3}) from the quadratic part of the action that defines the propagators.  The factor of $\omega $ in (\ref{VD4.3}) is included in order to show the gauge condition dependence present in the traditional effective action.  $\omega$ should be taken to be unity in the Vilkovisky-DeWitt result.  By keeping $\omega $ present we can compare the use of a de Donder (or harmonic) gauge ($\omega = 0 $) with the Landau-DeWitt gauge ($\omega = 1 $).

One important comment is that the formalism of the Vilkovisky-DeWitt effective action ensures that the results are independent of the choice made for $\omega $; however, this will not be shown in the present calculation because we have restricted attention to the Landau-DeWitt gauge for expediency.  If we wish to keep $\omega $ general, then the neglect of the terms in the connection denoted by $T^{k}_{ij}$ in (\ref{VD2.17}) is not justified; it is the presence of such terms that ensures the result for general $\omega $ agrees with that for $\omega = 1$ in the Landau-DeWitt gauge.  It should be possible to show this explicitly, although the calculations will be much more involved than those presented in the present paper and will be given elsewhere.

Another comment worth making is that we can use the gauge condition (\ref{VD4.3}) to set the $\partial^\mu a_\mu $ term in (\ref{VD4.2}) to zero.  (This is true because the gauge conditions appear as $\delta $-functions in the functional integral before they are promoted to exponentials.) We will do this later because it simplifies the calculations, although we will keep it present for the moment.  We have checked explicitly that the $\partial^\mu a_\mu $ term in (\ref{VD4.2}) makes no contribution to the electromagnetic field renormalization to verify this formal conclusion.

Because we have two gauge conditions we will have two terms arising from uncondensing (\ref{VD4.1}).  We will call the gauge parameters $\xi $ and $\zeta $ and take
\begin{equation}\label{VD4.7}
S_{GF}=\frac{1}{4\xi}\int d^nx\chi_\lambda^2+\frac{1}{4\zeta}\int d^nx\chi^2.
\end{equation}
The Landau-DeWitt gauge condition is specified by taking the $\xi\rightarrow0$ and $\zeta\rightarrow0$ limits. We will keep $\xi$ general to show the gauge condition dependence of the traditional background-field result, but take $\zeta\rightarrow0$ to simplify the expressions obtained.

After some calculation the result for $S_q$ in (\ref{VD3.29}) can be written as
\begin{equation}\label{VD4.8}
S_q=S_0+S_1+S_2,
\end{equation}
where
\begin{eqnarray}
S_0&=&\int d^nx\left\lbrace-\frac{1}{2}h^{\mu\nu}\Box h_{\mu\nu}+\frac{1}{4}h\Box h+\left(\frac{1}{\kappa^2\xi}-1\right)\left(\partial^\mu h_{\mu\nu}-\frac{1}{2}\partial_\nu h \right)^2\right.\nonumber\\&&\left.\qquad\qquad-\Lambda\left(h^{\mu\nu}h_{\mu\nu}-\frac{1}{2}h^2\right) \left\lbrack 1+\frac{v}{2}\left(\frac{n-4}{2-n}\right) \right\rbrack\right.\nonumber\\
&&\qquad\qquad\left.+\frac{1}{2}a_\mu\left(-\delta^{\mu\nu}\Box+\partial^\mu\partial^\nu\right)a_\nu +\frac{1}{4\zeta}\left(\partial^\mu a_\mu\right)^2 -\frac{v}{2}\Lambda\delta^{\mu\nu}a_\mu a_\nu\right\rbrace,\label{VD4.9}\\
S_1&=&\frac{\kappa}{2}\int d^nx\left(\bF^{\mu\nu}h\partial_\mu a_\nu-2 \bF_{\alpha}{}^{\nu}h^{\mu\alpha}\partial_\mu a_\nu +2 \bF_{\alpha}{}^{\nu}h^{\mu\alpha}\partial_\nu a_{\mu}\right)\nonumber\\ && -\frac{\kappa v}{4}\int d^nx\left( \delta^{\lambda}_{\sigma}\delta^{\mu\nu} -\delta^{\mu}_{\sigma}\delta^{\lambda\nu}-\delta^{\nu}_{\sigma}\delta^{\lambda\mu}  \right)\partial_\tau\bF^{\sigma\tau}h_{\mu\nu}a_\lambda\nonumber\\
&&+\frac{\omega}{\kappa\xi}\int d^nx\left(\partial^\mu h_{\mu\nu}-\frac{1}{2}\partial_\nu h\right)\left(\bar{A}^\nu\partial^\lambda a_\lambda+a^\lambda\bF_{\lambda}{}^{\nu}\right),\label{VD4.10}\\
S_2&=&\frac{\kappa^2}{4}\int d^nx\bF_{\mu\nu}\bF_{\alpha\beta}\left( 2\delta^{\mu\alpha}h^{\nu}_{\lambda}h^{\lambda\beta} +h^{\mu\alpha}h^{\nu\beta}-\delta^{\mu\alpha}hh^{\nu\beta}\right)\nonumber\\
&&-\frac{\kappa^2}{16}\left(1+\frac{v(n-4)}{4-2n}\right)\int d^nx \bF_{\alpha\beta}\bF^{\alpha\beta}\left(h^{\mu\nu}h_{\mu\nu}-\frac{1}{2}h^2\right) \nonumber\\
&&+\frac{\kappa^2 v}{4}\int d^nx \Big( -\bF^{\mu}{}_{\gamma}\bF^{\sigma\gamma}\delta^{\nu\lambda} + \frac{1}{4} \bF^{\lambda}{}_{\gamma}\bF^{\sigma\gamma}\delta^{\mu\nu} + \frac{1}{4}\bF^{\mu}{}_{\gamma}\bF^{\nu\gamma}\delta^{\lambda\sigma}-\frac{1}{2(2-n)} \delta^{\mu\lambda}\delta^{\sigma\nu}\bF_{\alpha\beta}\bF^{\alpha\beta}\nonumber\\
&& +\frac{1}{4(2-n)} \delta^{\mu\nu}\delta^{\lambda\sigma}\bF_{\alpha\beta}\bF^{\alpha\beta} \Big)h_{\mu\nu}h_{\lambda\sigma}-\frac{\kappa^2 v}{4}\int d^nx\left(\frac{1}{8}\delta^{\mu\nu}\bF_{\alpha\beta}\bF^{\alpha\beta}-\frac{1}{2}\bF^{\mu}{}_{\lambda}\bF^{\nu\lambda} \right)a_\mu a_\nu\nonumber\\
&& +\frac{\omega^2}{4\xi}\int d^nx\bF^{\mu}{}_{\lambda}\bF^{\nu\lambda}a_\mu a_\nu\label{VD4.11}
\end{eqnarray}

In these expressions the subscript $0, 1, 2 $ on $S$ denotes the order in the background gauge field $\bar{A}_\mu$ and we have shown explicitly the Vilkovisky-DeWitt terms with the  factor  $v$ as described above.  The traditional result is obtained using $v= 0$.  The spacetime dimension $n$ has been kept general at this stage, although we will be interested ultimately in the limit $n\rightarrow4$.  Because our concern here is only with pole terms in the effective action, it can be seen that as $n\rightarrow4$ all terms involving factors of $(n-4)$ , such as occur in (\ref{VD4.9}) and (\ref{VD4.11}), will not contribute.  Another observation is that the Vilkovisky-DeWitt connection leads to a term in $S_0$ that acts like a photon mass if $\Lambda\ne0$.

The  graviton and photon propagators follow from $S_0$ in the usual way.  The terms in $S_1$ and $S_2$ will be treated as interactions. We can write the photon propagator as
\begin{equation}\label{VD4.12}
G_{\mu\nu}(x,x')=\intp{p}e^{ip\cdot(x-x')}G_{\mu\nu}(p),
\end{equation}
and the graviton propagator as
\begin{equation}\label{VD4.13}
G_{\rho\sigma\lambda\tau}(x,x')=\intp{p}e^{ip\cdot(x-x')}G_{\rho\sigma\lambda\tau}(p).
\end{equation}
Using the result for $S_0$ leads to
\begin{equation}\label{VD4.14}
G_{\mu\nu}(p)=\frac{\delta_{\mu\nu}}{p^2-v\Lambda}+(2\zeta-1)\frac{p_\mu p_\nu}{(p^2-v\Lambda)(p^2-2\zeta v\Lambda)}\;.
\end{equation}
and,
\begin{equation}\label{VD4.15}
G_{\rho\sigma\lambda\tau}(p) =\frac{\delta_{\rho\lambda}\delta_{\sigma\tau}+\delta_{\rho\tau}\delta_{\sigma\lambda} -\frac{2}{n-2}\delta_{\rho\sigma}\delta_{\lambda\tau}}{2\left(p^2-2\lambda\right)}+\frac{1}{2}(\kappa^2\xi-1)\frac{\delta_{\rho\lambda}p_\sigma p_\tau+\delta_{\rho\tau}p_\sigma p_\lambda+\delta_{\sigma\lambda}p_\rho p_\tau+\delta_{\sigma\tau}p_\rho p_\lambda}{\left(p^2-2\lambda\right) \left(p^2-2\kappa^2\xi\lambda\right)},
\end{equation}
where we have defined
\begin{equation}\label{VD4.16}
\lambda=\Lambda+v\Lambda\left(\frac{n-4}{4-2n}\right).
\end{equation}
In our calculations of the pole terms, the Vilkovisky-DeWitt correction in (\ref{VD4.16}) will make no contributions to the poles when $n\rightarrow4$, and we may set $\lambda\rightarrow\Lambda$ in this limit.  This will not be true for the finite part of the effective action or in spacetimes of dimension other than four.  (Of course the Vilkovisky-DeWitt correction enters the calculation in other places through the interaction terms in any case.)

As explained we will treat the terms $S_1+S_2$ as an interaction.  Simple power counting shows that the divergent part of the effective action can involve $\bF_{\mu\nu}$ up to and including terms of fourth order.  (In more than four spacetime dimensions, higher powers of $\bF_{\mu\nu}$ must be considered.) We can write
\begin{equation}\label{VD4.17}
\Gamma_G=\langle e^{-S_1-S_2}-1\rangle
\end{equation}
where $\langle\cdots\rangle$ means to evaluate the enclosed expression using Wick's theorem and the basic pairings
\begin{eqnarray}
\langle a_\mu(x)a_\nu(x')\rangle&=&G_{\mu\nu}(x,x'),\label{VD4.18}\\
\langle h_{\rho\sigma}(x)h_{\lambda\tau}(x')\rangle&=&G_{\rho\sigma\lambda\tau}(x,x'),\label{VD4.19}
\end{eqnarray}
If we drop terms of order $\bF^5$ and higher, use of Wick's theorem shows that
\begin{equation}\label{VD4.20}
\Gamma_G=\Gamma_{G2}+\Gamma_{G4}+\cdots,
\end{equation}
where $ \Gamma_{Gk}$ is of order, $\bF^k$.  There is no cubic term in $\bF$ present as claimed earlier because such terms can only arise from those in the expansion of $\Gamma_G $ that involve odd numbers of graviton and photon fields; these vanish upon use of Wick's theorem. ({\em e.g.\/} $\langle S_1S_2\rangle=0=\langle S_1^3\rangle$.)  We find
\begin{eqnarray}
\Gamma_{G2}&=&\langle S_2\rangle-\frac{1}{2}\langle S_1^2\rangle,\label{VD4.21}\\
\Gamma_{G4}&=&-\frac{1}{2}\langle S_2^2\rangle+\frac{1}{2}\langle S_1^2S_2\rangle-\frac{1}{24}\langle S_1^4\rangle.\label{VD4.21b}
\end{eqnarray}
The (correct) Vilkovisky-DeWitt result is obtained by taking the parameters $v=1,\omega=1,\xi=0,\zeta=0$ in these expressions.  We now examine these two terms separately, and then turn to the possible ghost contributions.

\subsection{Evaluation of $\Gamma_{G2}$}\label{G2}

We first of all use Wick's theorem to evaluate both $\langle S_2\rangle$ and $\langle S_1^2\rangle$.  This will give us the results in terms of the graviton and photon propagators.  The momentum space representations (\ref{VD4.12}--\ref{VD4.15}) can be used and the resulting integrals evaluated using standard methods.  (We give the basic results in the appendix.)

For $\langle S_2\rangle$ we find, after use of Wick's theorem,
\begin{equation}\label{VD4.22}
\langle S_2\rangle=\langle S_{21}\rangle+\langle S_{22}\rangle,
\end{equation}
where
\begin{eqnarray}
\langle S_{21}\rangle&=&\frac{\kappa^2}{4}\int d^nx\Big\lbrace(2-v)\Big(\bF^{\mu}{}_{\gamma}\bF^{\beta\gamma}\delta^{\nu\alpha} -\frac{1}{2}\delta^{\mu\nu}\bF^{\alpha}{}_{\gamma}\bF^{\beta\gamma}\Big) +\bF^{\mu\alpha}\bF^{\nu\beta}\nonumber\\
&&\hspace{60pt}+\frac{1}{4}(v-1)\bF^2\Big(\delta^{\mu\alpha}\delta^{\nu\beta} -\frac{1}{2}\delta^{\mu\nu}\delta^{\alpha\beta}\Big)\Big\rbrace G_{\mu\nu\alpha\beta}(x,x),\label{VD4.23}\\
\langle S_{22}\rangle&=&\frac{\kappa^2}{4}\int d^nx\Big\lbrace \frac{v}{2}\Big(\bF^{\mu}{}_{\gamma}\bF^{\nu\gamma}-\frac{1}{4}\bF^2\delta^{\mu\nu}\Big) +\frac{\omega^2}{\kappa^2\xi}\,\bF^{\mu}{}_{\gamma}\bF^{\nu\gamma}\Big\rbrace G_{\mu\nu}(x,x). \label{VD4.24}
\end{eqnarray}
We have abbreviated $\bF^2=\bF_{\mu\nu}\bF^{\mu\nu}$ here and in the following.  The two terms (\ref{VD4.23}) and (\ref{VD4.24}) involve only the coincidence limit of the Green functions and no derivatives of the electromagnetic field strength $\bF$.  Any pole terms will contribute to the charge renormalization.

We use dimensional regularization with only the logarithmic divergences present as described earlier.  Because we are only interested in pole terms of the effective action coming from logarithmic divergences we can adopt the method described in the appendix.  The results, after some calculation, turn out to be given by (where $L$ stands for the basic logarithmic divergence defined in (\ref{A11},\ref{A13}))
\begin{eqnarray}
\langle S_{21}\rangle&=&\frac{3}{8}\kappa^2\Lambda(\kappa^4\xi^2+1)L\int d^4x\bF^2,\label{VD4.25}\\
\langle S_{22}\rangle&=&\frac{\omega^2}{4\xi}v\Lambda L\int d^4x\bF^2.\label{VD4.26}
\end{eqnarray}
We have let $\zeta\rightarrow0 $ in (\ref{VD4.26}).  This is essential because we have dropped the $\partial^\lambda a_\lambda$ term in $S_1$ to shorten the expressions obtained and simplify the calculation and it would be inconsistent to retain $\zeta$.  The first term (the one multiplied by $v$) of $\langle S_{22}\rangle$ in (\ref{VD4.24}) does not contribute to the result since $G_{\mu\nu}(x,x)\propto\delta_{\mu\nu}$ and the contraction of this with the field strength vanishes.  (This can be recognized as involving the trace of the stress-energy-momentum tensor for the electromagnetic field which vanishes for $n=4$.) The Vilkovisky-DeWitt parameter does not enter (\ref{VD4.25}) although this was not obvious from (\ref{VD4.23}).  Combining the two results (\ref{VD4.25}) and (\ref{VD4.26}) results in
\begin{equation}\label{VD4.27}
\langle S_{2}\rangle=\frac{3}{8}\kappa^2\left(1+\kappa^4\xi^2+\frac{\omega^2 v }{2\kappa^2\xi}\right)\Lambda L\int d^4x\bF^2,
\end{equation}
as the relevant pole part.  For the Vilkovisky-DeWitt result we take $\omega=v=1$ and try to let $\xi\rightarrow0$.  However, there is a term in $1/\xi$ present that prohibits this limit to be taken completely.  Because the Vilkovisky-DeWitt formalism ensures that $\xi\rightarrow0$ must exist and be finite, cancellation of all terms that are singular in this limit and that appear at intermediate stages of the calculation provides a useful check on the results.

In order to evaluate $\langle S_{1}^2\rangle$ (and the higher order terms in $\Gamma_{G4}$) it proves convenient to write $S_1$ in (\ref{VD4.10}) as
\begin{equation}\label{VD4.28}
S_1=\int d^nx\left( P^{\alpha\beta\mu\nu}h_{\alpha\beta}\partial_\mu a_\nu+P^{\alpha\beta\lambda}h_{\alpha\beta}a_\lambda\right),
\end{equation}
where
\begin{eqnarray}
P^{\alpha\beta\mu\nu}&=&\frac{\kappa}{2}\left\lbrace \Big(1-\frac{\omega}{\kappa^2\xi}\Big)\left\lbrack \bF^{\mu\nu}\delta^{\alpha\beta}-\bF^{\beta\nu}\delta^{\mu\alpha} -\bF^{\alpha\nu}\delta^{\mu\beta}\right\rbrack
+\bF^{\beta\mu}\delta^{\nu\alpha} +\bF^{\alpha\mu}\delta^{\nu\beta}\right\rbrace,\label{VD4.29}\\
P^{\alpha\beta\lambda}&=&\frac{\kappa v}{4}
\left( \delta^{\lambda\alpha}\partial_\tau\bF^{\beta\tau} +\delta^{\lambda\beta}\partial_\tau\bF^{\alpha\tau}\right)-\frac{\omega}{2\kappa\xi}(\partial^\alpha\bF^{\lambda\beta} +\partial^\beta\bF^{\lambda\alpha})\nonumber\\
&&\quad +\left(\frac{\omega}{2\kappa\xi}
-\frac{\kappa v}{4}\right)\delta^{\alpha\beta}\partial_\tau\bF^{\lambda\tau}.\label{VD4.30}
\end{eqnarray}

Wick's theorem gives us
\begin{eqnarray}
\langle S_1^2\rangle&=&\int d^nx\int d^nx'\Big\lbrace P^{\alpha\beta\mu\nu}(x)P^{\lambda\sigma\gamma\delta}(x') G_{\alpha\beta\lambda\sigma}(x,x') \partial_{\mu}\partial_{\gamma}^{\prime}G_{\nu\delta}(x,x')\nonumber\\
&&+2P^{\alpha\beta\mu\nu}(x)P^{\lambda\sigma\gamma}(x') G_{\alpha\beta\lambda\sigma}(x,x') \partial_{\mu}G_{\nu\gamma}(x,x')\nonumber\\
&&+P^{\alpha\beta\gamma}(x)P^{\lambda\sigma\delta}(x') G_{\alpha\beta\lambda\sigma}(x,x') G_{\nu\delta}(x,x')\Big\rbrace.\label{VD4.31}
\end{eqnarray}
The products of Green functions may be evaluated using the momentum space representations and results of the Appendix. After considerable calculation it may be shown that
\begin{equation}\label{VD4.32}
\langle S_1^2\rangle=\kappa^2\alpha_1 L\int d^4x\bF^2+\kappa^2\beta L\int d^4x\left(\partial^\mu\bF_{\mu\nu}\right)^2,
\end{equation}
where
\begin{eqnarray}
\alpha_1&=&\frac{3\Lambda v\omega^2}{8\kappa^2\xi}-\frac{3}{4}\Lambda v\omega +\frac{3}{8}\Lambda v+\frac{3}{4}\Lambda\omega^2+\frac{3}{4}\Lambda\nonumber\\
&&+\frac{3}{8}\Lambda v\kappa^2\xi-\frac{3}{2}\Lambda\omega\kappa^2\xi+\frac{3}{4}\Lambda\kappa^4\xi^2,\label{VD4.33}\\
\beta&=&-\frac{1}{12}+\frac{2}{3}\omega+\frac{3}{16}v^2 +\frac{1}{4}v+\frac{1}{2}v\omega+\frac{1}{4}\kappa^2\xi\nonumber\\
&&+\frac{3}{16}v^2\kappa^2\xi-v\kappa^2\xi.\label{VD4.34}
\end{eqnarray}
In writing down this expression we have chosen to write the term that involves derivatives of $\bF$ as shown. In the calculation we also find a term $\bF_{\mu\nu}\Box\bF^{\mu\nu}$ that when integrated by parts is equivalent to $-2(\partial^\mu\bF_{\mu\nu})^2$.

We can now find the complete pole part of the effective action that is quadratic in $\bF$ and comes from the gauge field and graviton. (We still need to find the ghost contribution and we will do this in the next section.) From (\ref{VD4.21},\ref{VD4.27},\ref{VD4.32}--\ref{VD4.34}) we have
\begin{equation}\label{VD4.35}
\Gamma_{G2}=\kappa^2\alpha \Lambda L\int d^4x\bF^2-\frac{1}{2}\kappa^2\beta L\int d^4x\left(\partial^\mu\bF_{\mu\nu}\right)^2,
\end{equation}
where
\begin{equation}\label{VD4.36}
\alpha=\frac{3}{8}v\omega-\frac{3}{16}v -\frac{3}{8}\omega^2-\frac{3}{16}v\kappa^2\xi +\frac{3}{4}\omega\kappa^2\xi,
\end{equation}
and $\beta$ is given in (\ref{VD4.34}).

There are several comments to be made at this stage. The first is that although terms that are singular as $\xi\rightarrow0$ occur at intermediate stages of the calculation (see (\ref{VD4.27}) for example), all such terms cancel when we form the effective action as guaranteed by the general formalism. A second comment is that the coefficients of both terms in (\ref{VD4.35}) depend on the choice of gauge condition parameters $\omega$ and $\xi$ even if we take the Vilkovisky-DeWitt parameter $v=0$ corresponding to the use of the standard background-field method. Unless special care is taken when using the traditional background-field method, or equivalently the naive Feynman rules, results will be obtained for the effective action that are gauge condition dependent. This is completely obscured in calculations that fix any of these parameters at the start for calculational convenience. A final comment is that if the cosmological constant vanishes then there is no contribution to the term in $\bF^2$ that is responsible for the electromagnetic field, and hence the charge renormalization, in agreement with earlier results of \cite{Piet,DJT1,Ebertetal}. The result for $\Lambda\ne0$ was first given in \cite{DJT2}.

\subsection{Evaluation of $\Gamma_{G4}$}\label{G4}

We begin with each of the three terms that comprise the contributions to $\Gamma_{G4}$. On dimensional grounds there can be no derivatives of the background electromagnetic field, so we may safely take $\bF_{\mu\nu}$ to be constant. This simplifies the calculation. There are two independent invariants that are gauge invariant and we take them to be $(\bF^2)^2$ and $\bF^4$ where
\begin{eqnarray}
\bF^2&=&\bF_{\mu\nu}\bF^{\mu\nu},\label{G4.1a}\\
\bF^4&=&\bF_{\mu\nu}\bF^{\nu\lambda}\bF_{\lambda\sigma}\bF^{\sigma\mu}.\label{G4.1b}
\end{eqnarray}
We will write
\begin{equation}\label{G4.1}
\Gamma_{G4}=\kappa^4 L\int d^4x\left\lbrace A (\bF^2)^2 + B\bF^4\right\rbrace,
\end{equation}
for some coefficients $A$ and $B$. Nether $A$ nor $B$ can depend on the cosmological constant (on dimensional grounds); thus, the result that we will obtain for the pole part of the effective action indicated in (\ref{G4.1}) will apply equally well to Einstein-Maxwell theory without a cosmological constant.

We begin by noting that the term called $P^{\alpha\beta\lambda}$ in (\ref{VD4.28},\ref{VD4.30}) cannot contribute to the pole terms in (\ref{G4.1}) as it vanishes when we set $\bF_{\mu\nu}$ to be constant. We can write $S_2$ in (\ref{VD4.11}) in a convenient way as
\begin{equation}\label{G4.2}
S_2=\int d^nx\left( R^{\mu\nu\alpha\beta}h_{\mu\nu}h_{\alpha\beta}+R^{\mu\nu}a_{\mu}a_{\nu}\right),
\end{equation}
where $R^{\mu\nu\alpha\beta}$ and $R^{\mu\nu}$ can be read off by comparison of (\ref{G4.2}) with (\ref{VD4.11}) and the results symmetrized in the obvious way. Both $R^{\mu\nu\alpha\beta}$ and $R^{\mu\nu}$ may be taken to be constant for our purposes. Application of Wick's theorem gives
\begin{eqnarray}
\left\langle (S_2)^2\right\rangle&=&2\int d^nx d^nx'\Big\lbrace R^{\mu\nu\alpha\beta}R^{\rho\sigma\lambda\tau} G_{\mu\nu\rho\sigma}(x,x')G_{\alpha\beta\lambda\tau}(x,x')\nonumber\\
&&\quad + R^{\mu\nu}R^{\rho\sigma} G_{\mu\rho}(x,x')G_{\nu\sigma}(x,x')\Big\rbrace.\label{G4.3}
\end{eqnarray}
The products of Green functions are evaluated as described in the Appendix and the results are then contracted with $R^{\mu\nu\alpha\beta}$ and $R^{\mu\nu}$ in (\ref{G4.3}). The result takes the form on the right hand side of (\ref{G4.1}) where $A=A_1$ and $B=B_1$ with
\begin{eqnarray}
A_1&=&\frac{\omega^4}{192\kappa^4\xi^2}-\frac{7v\omega^2}{384\kappa^2\xi} +\frac{v}{32}-\frac{23 v^2}{1536}+\frac{1}{64},\label{G4.4}\\
B_1&=&\frac{7\omega^4}{96\kappa^4\xi^2}+\frac{7v\omega^2}{96\kappa^2\xi} -\frac{v}{8}+\frac{23 v^2}{384}+\frac{1}{32}.\label{G4.5}
\end{eqnarray}

The next term of order $\bF^4$ involves
\begin{eqnarray}
\left\langle (S_1)^2S_2\right\rangle&=&2\int d^nx d^nx' d^nx''P^{\alpha\beta\mu\nu}P^{\lambda\sigma\gamma\delta}\Big\lbrace R^{\epsilon\rho\theta\phi} \partial_\mu\partial_{\gamma}^{\prime} G_{\nu\delta}(x,x')G_{\alpha\beta\epsilon\rho}(x,x'') G_{\lambda\sigma\theta\phi}(x',x'')\nonumber\\
&&+R^{\epsilon\rho}G_{\alpha\beta\lambda\sigma}(x,x')\partial_{\mu}G_{\nu\epsilon}(x,x'')\partial_{\gamma}^{\prime}G_{\delta\rho}(x',x'')\Big\rbrace.\label{G4.6}
\end{eqnarray}
The products of Green functions are evaluated as before, and we again find a result taking the form on the right hand side of (\ref{G4.1}) where this time $A=A_2$ and $B=B_2$ with
\begin{eqnarray}
A_2&=&\frac{\omega^4}{96\kappa^4\xi^2}-\frac{7v\omega^2}{384\kappa^2\xi}+ \frac{5\omega^2}{96\kappa^2\xi}-\frac{\omega^3}{48\kappa^2\xi} - \frac{23v\kappa^2\xi}{384}+\frac{5v\omega}{64}+\frac{3 v}{128}+\frac{5\omega^2}{192}+\frac{3\kappa^2\xi}{16}\nonumber\\
&&\quad-\frac{\omega\kappa^2\xi}{32} +\frac{v\kappa^4\xi^2}{48}-\frac{v\omega\kappa^2\xi}{24}+\frac{v\omega^2}{48} -\frac{3\omega}{16} +\frac{\kappa^4\xi^2}{64}+\frac{3}{64} ,\label{G4.7}\\
B_2&=&\frac{7\omega^4}{48\kappa^4\xi^2}+\frac{7v\omega^2}{96\kappa^2\xi}- \frac{\omega^2}{48\kappa^2\xi}-\frac{7\omega^3}{24\kappa^2\xi} + \frac{23v\kappa^2\xi}{96}-\frac{5v\omega}{16}-\frac{3 v}{32}+\frac{13\omega^2}{48}-\frac{3\kappa^2\xi}{8}\nonumber\\
&&\quad-\frac{\omega\kappa^2\xi}{4} -\frac{v\kappa^4\xi^2}{12}+\frac{v\omega\kappa^2\xi}{6}-\frac{v\omega^2}{12} +\frac{3\omega}{8} +\frac{\kappa^4\xi^2}{8}.\label{G4.8}
\end{eqnarray}

The third and final piece of $\Gamma_{G4}$ involves $\langle (S_1)^4\rangle$. The Wick reduction leads to
\begin{eqnarray}
\left\langle (S_1)^4\right\rangle&=&3\int d^nxd^nx'd^nx''d^nx'''P^{\alpha\beta\mu\nu} P^{\lambda\sigma\rho\delta} P^{\theta\phi\psi\chi} P^{\kappa\tau\epsilon\iota} G_{\alpha\beta\lambda\sigma}(x,x') G_{\theta\phi\kappa\tau}(x'',x''')\nonumber\\
&&\times\left\lbrack \partial_\mu\partial_{\psi}^{\prime\prime}G_{\nu\chi}(x,x'') \partial_{\rho}^{\prime}\partial_{\epsilon}^{\prime\prime\prime}G_{\delta\iota}(x',x''') +\partial_\mu\partial_{\epsilon}^{\prime\prime\prime}G_{\nu\iota}(x,x''') \partial_{\rho}^{\prime}\partial_{\psi}^{\prime\prime}G_{\delta\chi}(x',x'')\right\rbrack.\label{G4.9a}
\end{eqnarray}
Evaluating the products of Green functions leads to a result taking the form on the right hand side of (\ref{G4.1}) where $A=A_3$ and $B=B_3$ with
\begin{eqnarray}
A_3&=&\frac{\omega^4}{16\kappa^4\xi^2}+\frac{5\omega^2}{8\kappa^2\xi} -\frac{\omega^3}{4\kappa^2\xi} + \frac{3\omega^2}{8}
-\frac{\omega\kappa^2\xi}{4} +\frac{\kappa^4\xi^2}{16}-\frac{5\omega}{4}+\frac{5\kappa^2\xi}{8} +\frac{3}{16},\label{G4.9}\\
B_3&=&\frac{7\omega^4}{8\kappa^4\xi^2}-\frac{\omega^2}{4\kappa^2\xi} -\frac{7\omega^3}{2\kappa^2\xi} + \frac{21\omega^2}{4}
-\frac{7\omega\kappa^2\xi}{2} +\frac{7\kappa^4\xi^2}{8}+\frac{\omega}{2}-\frac{\kappa^2\xi}{4} +\frac{3}{8}.\label{G4.10}
\end{eqnarray}

We have kept $\omega,\ v$ and $\xi$ present to demonstrate that individual terms are singular as $\xi\rightarrow0$, and that the results computed using the standard background-field method are gauge condition dependent. The net result for $\Gamma_{G4}$ follows as (\ref{G4.1}) with
\begin{eqnarray}
A&=&-\frac{1}{2}A_1+\frac{1}{2}A_2-\frac{1}{24}A_3\nonumber\\
&=&-\frac{\omega^2}{384}-\frac{\omega}{24}+\frac{1}{128}+\frac{13\kappa^2\xi}{192}-\frac{\omega\kappa^2\xi}{192}+\frac{\kappa^4\xi^2}{192}\nonumber\\
&&+v\left(\frac{\omega^2}{96}-\frac{23\kappa^2\xi}{768} +\frac{23v}{3072} +\frac{5\omega}{128}+\frac{\kappa^4\xi^2}{96}- \frac{\omega\kappa^2\xi}{48}-\frac{1}{256}\right),\label{G4.11}\\
B&=&-\frac{1}{2}B_1+\frac{1}{2}B_2-\frac{1}{24}B_3\nonumber\\
&=&-\frac{\omega^2}{12}+\frac{\omega}{6}-\frac{1}{32}-\frac{17\kappa^2\xi}{96} +\frac{\omega\kappa^2\xi}{48}+\frac{5\kappa^4\xi^2}{192}\nonumber\\
&&+v\left(-\frac{\omega^2}{24}+\frac{23\kappa^2\xi}{192}- \frac{23 v}{768} -\frac{5\omega}{32}-\frac{\kappa^4\xi^2}{24}+ \frac{\omega\kappa^2\xi}{12}+\frac{1}{64}\right).\label{G4.12}
\end{eqnarray}
As with our earlier calculation, individual contributions to the effective action contain singular terms as $\xi\rightarrow0$; however, when all terms of the same order are combined all such singular behaviour cancels to leave a finite result as $\xi\rightarrow0$. We again see that if $v=0$, the traditional result for the effective action is gauge dependent. The correct, gauge condition independent result can be found from $\omega=v=1$ and $\xi=0$. There is still the ghost contribution to consider and this is the subject of the next section.

\section{Ghost contribution}\label{VD5}
\subsection{Expansion of the effective action}\label{ghost1}

We can evaluate the ghost contribution to the effective action in the same way as we did for the graviton and gauge fields.  From (\ref{VD3.30}) we identify the ghost action as
\begin{equation}\label{VD5.1}
S_{GH}=\bar{\eta}_\alpha Q^{\alpha}{}_{\beta}\eta^\beta,
\end{equation}
with $ Q^{\alpha}{}_{\beta}$ given by (\ref{VD2.5}).  It can be noted that
\begin{equation}\label{VD5.2}
Q^{\alpha}{}_{\beta}\eta^\beta= \chi^{\alpha}{}_{,i}K^{i}_{\beta}\eta^\beta=\delta\chi^{\alpha},
\end{equation}
where $\delta\chi^\alpha$ represents the change in the gauge condition under a gauge transformation with the infinitesimal gauge parameters $\delta\epsilon^\beta $ replaced with the anticommuting ghost field $\eta^\beta $.  The background fields are held fixed when computing $ Q^{\alpha}{}_{\beta}$.

In our case we have the two gauge conditions (\ref{VD4.2}) and (\ref{VD4.3}).  We need a vector ghost $\eta^\mu(x)$ and its antighost $\bar{\eta}_\mu(x)$ for gravity, and a scalar ghost $\eta (x)$ and its antighost $\bar{\eta}(x) $ for electromagnetism.  The ghost action will be
\begin{equation}\label{VD5.3}
S_{GH}=\int d^nx\left(\bar{\eta}^\lambda\delta\chi_\lambda+\bar{\eta}\delta\chi\right).
\end{equation}
Here, $\delta\chi_\lambda $ and $\delta\chi $ denote the changes in the gauge conditions (\ref{VD4.2}) and (\ref{VD4.3}) under a gauge transformation of the metric and electromagnetic field ((\ref{VD3.7}) and (\ref{VD3.8})) using (\ref{VD4.5}) and (\ref{VD4.6}) with the gauge parameters $\delta\epsilon^\lambda(x)\rightarrow\eta^\lambda (x) $ and $\delta\epsilon (x)\rightarrow\eta (x) $.  Furthermore, because we are only working to one-loop order we can neglect all terms in $S_{GH}$ that involve the quantum fields $h_{\mu\nu} $ and $a_\mu $.  (They would be important at higher loop orders.)

The result for $S_{GH}$ can be conveniently expressed as a sum of three terms,
\begin{equation}\label{VD5.4}
S_{GH}=S_{GH0}+S_{GH1}+S_{GH2},
\end{equation}
with the subscript $0,1,2$ counting the power of the background gauge field that occurs just as we did earlier.  We have
\begin{eqnarray}
S_{GH0}&=&\int d^nx\left(-\frac{2}{\kappa^2}\bar{\eta}^\lambda\Box\eta_\lambda -\bar{\eta}\Box\eta\right),\label{VD5.5}\\
S_{GH1}&=&\int d^nx\vphantom{|}\left\lbrack\omega\bar{\eta}^\lambda\bF_{\mu\lambda}\eta^{,\mu} +\bar{\eta}\left(\bar{A}_{\mu,\nu}+\bar{A}_{\nu,\mu}\right)\eta^{\nu,\mu}\right.\nonumber\\
&&\vphantom{|}\left.\quad + \bar{\eta} \bar{A}^{\mu}_{,\mu\nu}\eta^\nu +\bar{\eta}\bar{A}_\nu\Box\eta^\nu\right\rbrack,\label{VD5.6}\\
S_{GH2}&=&\omega \int d^nx\, \bar{\eta}^\lambda\bF_{\mu\lambda} \left( -\bar{A}^{\mu}_{,\nu}\eta^\nu - \bar{A}_\nu \eta^{\nu,\mu}\right).\label{VD5.7}
\end{eqnarray}

We can again treat the terms that involve the background gauge field $S_{GH1}$ and $S_{GH2}$ as interaction terms and in place of (\ref{VD4.17}) have
\begin{equation}\label{VD5.8}
\Gamma_{GH}=-\langle e^{-S_{GH1}-S_{GH2}}-1\rangle,
\end{equation}
with the overall minus sign due to the ghost statistics.  We have the basic pairing relations
\begin{eqnarray}
\langle\eta_{\mu}(x)\bar{\eta}_\nu(x')\rangle&=&\Delta_{\mu\nu}(x,x'),\label{VD5.9}\\
\langle\eta(x)\bar{\eta}(x')\rangle&=&\Delta(x,x'),\label{VD5.10}
\end{eqnarray}
where
\begin{equation}\label{VD5.11}
\Delta_{\mu\nu}(x,x')=\frac{\kappa^2}{2}\delta_{\mu\nu}\Delta(x,x'),
\end{equation}
and,
\begin{equation}\label{VD5.12}
-\Box\Delta(x,x')=\delta(x,x'),
\end{equation}
follow from (\ref{VD5.5}).

We find, up to fourth order in the background gauge field
\begin{eqnarray}
\Gamma_{GH}&=&-\langle S_{GH2}\rangle+\frac{1}{2}\langle (S_{GH1})^2\rangle+\frac{1}{2}\langle(S_{GH2})^2\rangle\nonumber\\
&&-\frac{1}{2}\langle(S_{GH1})^2S_{GH2}\rangle +\frac{1}{24}\langle(S_{GH1})^4\rangle.\label{VD5.13}
\end{eqnarray}
Again, the potential cubic terms in the background gauge field do not contribute because they involve odd numbers of ghost fields and vanish by application of the Wick reduction.

\subsection{Evaluation of $\Gamma_{GH2}$}\label{ghost2}
It is convenient to use the gauge condition for the electromagnetic field to simplify $\chi_\lambda$ in (\ref{VD4.2}). We can set the term in $\partial^\mu a_\mu$ in (\ref{VD4.2}) to zero as before, and this simplifies the evaluation of the ghost contributions. We have checked this by not making this simplification and replacing the second term of (\ref{VD4.2}) by $\omega'\bar{A}_\lambda\partial^\mu a_\mu+\omega a^\mu\bF_{\mu\lambda}$. It can then be shown that $\omega'$ cancels out of $\Gamma_{G2}$ and therefore may be safely taken to vanish without any loss of generality.

We find from $S_{GH2}$ in (\ref{VD5.7}) that
\begin{eqnarray}
\langle S_{GH2}\rangle&=&\omega\int d^nx\bF_{\mu\lambda}\left\lbrack \bar{A}^{\mu}_{,\nu} \Delta^{\nu\lambda}(x,x)+\bar{A}_\nu\left.\partial^\mu \Delta^{\nu \lambda}(x,x')\right|_{x'=x}\right\rbrack\nonumber\\
&=&0, \label{VD5.14}
\end{eqnarray}
since the coincidence limit of the Green functions involve massless propagators that get regularized to zero in dimensional regularization.

For $\langle(S_{GH1})^2\rangle$ we find
\begin{eqnarray}
\left\langle (S_{GH1})^2 \right\rangle&=& -2\omega\int d^nxd^nx'\Big\lbrace \bar{A}_{\mu,\nu}(x')\partial^{\prime\mu}\Delta^{\nu\lambda}(x',x) +\bar{A}^{\mu}_{,\mu\nu}(x')\Delta^{\nu\lambda}(x',x)\nonumber\\
&&+\bar{A}_{\nu}(x')\Box^{\prime}\Delta^{\nu\lambda}(x',x) +\bar{A}_{\nu,\mu}(x')\partial^{\prime\mu}\Delta^{\nu\lambda}(x',x)\Big\rbrace\bF_{\sigma\lambda}(x)\partial^\sigma \Delta(x,x'),\label{VD5.15a}
\end{eqnarray}
after Wick reduction using the pairing relations (\ref{VD5.9}--\ref{VD5.11}) with the ghosts treated as anticommuting. Evaluating the products of Green functions as before followed by some integration by parts, results in
\begin{equation}\label{VD5.15}
\left\langle (S_{GH1})^2 \right\rangle=-\frac{1}{6}\kappa^2\omega L\int d^4x\left(\partial^\mu\bF_{\mu\nu}\right)^2.
\end{equation}
Although separate terms of (\ref{VD5.15a}) are not gauge invariant, the net result is that all terms when combined lead to a gauge invariant answer. This is as it must be since the formalism guarantees that this is so. We therefore find
\begin{equation}\label{VD5.16}
\Gamma_{GH2}=-\frac{1}{12}\kappa^2\omega L\int d^4x \left(\partial^\mu\bF_{\mu\nu}\right)^2.
\end{equation}
This vanishes for $\bF_{\mu\nu}$ constant, so the ghosts make no contribution to the charge renormalization. They do however contribute to the pole part of the effective action for general background fields.

\subsection{Evaluation of $\Gamma_{GH4}$}\label{ghost4}

The Wick reduction of $\langle(S_{GH2})^2\rangle$ results in
\begin{eqnarray}
\left\langle(S_{GH2})^2\right\rangle&=&-\frac{\kappa^4\omega^2}{4}\int d^nxd^nx'\bF_{\mu\lambda}(x)\bF_{\alpha}{}^{\nu}(x')\Big\lbrace \bar{A}^{\mu}{}_{,\nu}(x)\bar{A}^{\alpha,\lambda}(x')\Delta(x,x')\Delta(x',x)\nonumber\\
&&+\bar{A}^{\mu}{}_{,\nu}(x)\bar{A}^{\lambda}(x')\Delta(x,x') \partial^{\prime\alpha}\Delta(x',x) +\bar{A}_{\nu}(x)\bar{A}^{\alpha,\lambda}(x') \partial^{\mu}\Delta(x,x')\Delta(x',x)\nonumber\\
&& +\bar{A}_{\nu}(x)\bar{A}^{\lambda}(x') \partial^{\mu}\Delta(x,x')\partial^{\prime\alpha}\Delta(x',x)\Big\rbrace.\label{GH4.1}
\end{eqnarray}
The calculations of the ghost contribution starts to become extremely messy if we keep the background gauge field $\bar{A}_\mu$ general. However we can simplify things enormously by noting that the final result must be expressible in terms of the two invariants $(\bF^2)^2$ and $\bF^4$ as we had earlier in (\ref{G4.1}). Because the result must be invariant under gauge transformations of the background field, we may simplify with the choice
\begin{equation}\label{GH4.2a}
\bar{A}_\mu(x)=-\frac{1}{2}\bF_{\mu\nu}x^\nu,
\end{equation}
so that $\partial_\mu\bar{A}_{\nu}=\frac{1}{2}\bF_{\mu\nu}$ and $\partial^\mu\bar{A}_\mu=0$. In order to implement this it is easiest to assume that $\bar{F}_{\mu\nu}$ is constant, and integrate by parts so that all derivatives act on factors of $\bar{A}_\mu$. After some work it can be shown that
\begin{equation}\label{GH4.2}
\left\langle(S_{GH2})^2\right\rangle=-\frac{7\kappa^4\omega^2}{64}L\int d^4x\bF^4.
\end{equation}

For $\langle(S_{GH1})^2S_{GH2}\rangle$ we find, after Wick reduction and use of the pairings (\ref{VD5.9}--\ref{VD5.11}),
\begin{eqnarray}
\left\langle(S_{GH1})^2S_{GH2}\right\rangle&=&\frac{1}{2}\kappa^4\omega^2\int d^nxd^nx'd^nx''\bF_{\mu\nu}(x)\partial^\mu\Delta(x,x')Y^{\lambda}(x')\Delta(x',x'')\nonumber\\
&&\times\bF_{\beta\lambda}(x'') \left\lbrack\partial^{\prime\prime\nu}\bar{A}^{\beta}(x'')+ \bar{A}^{\nu}(x'')\partial^{\prime\prime\beta}\right\rbrack\Delta(x'',x) ,\label{GH4.3}
\end{eqnarray}
where we have defined
\begin{equation}\label{GH4.4}
Y^\lambda(x)=\partial^\lambda\partial^\sigma\bar{A}_\sigma +\left(\partial^\lambda\bar{A}^\sigma +\partial^\sigma\bar{A}^\lambda\right)\partial_\sigma +\bar{A}^\lambda\Box.
\end{equation}
The result in (\ref{GH4.3}) can be evaluated as we described above for $\langle(S_{GH2})^2\rangle$ and the result turns out to be
\begin{equation}\label{GH4.5}
\left\langle(S_{GH1})^2S_{GH2}\right\rangle=\kappa^4\omega^2 L\int d^4x\left\lbrack \frac{1}{96}(\bF^2)^2-\frac{1}{48}\bF^4\right\rbrack.
\end{equation}

Finally we come to $\langle(S_{GH1})^4\rangle$ that proved to be the most lengthy to evaluate. The Wick reduction yields
\begin{eqnarray}
\left\langle(S_{GH1})^4\right\rangle&=&-3\kappa^4\omega^2\int d^nxd^nx'd^nx''d^nx''' \bF_{\mu\nu}(x)\bF_{\alpha\beta}(x') \partial^\mu\Delta(x,x'')\partial^{\prime\alpha}\Delta(x',x''') \nonumber\\
&&\quad\times Y^\beta(x'')\Delta(x'',x')Y^\nu(x''')\Delta(x''',x)\label{GH4.6}\\
&=&-\kappa^4\omega^2 L\int d^4x\left\lbrack\frac{1}{4}(\bF^2)^2+\frac{3}{16}\bF^4 \right\rbrack,
\end{eqnarray}
with the second equality following after some calculation.

We can now form the complete ghost contribution to the effective action that is quartic in the background gauge field from the last three terms of (\ref{VD5.13}). The result is
\begin{equation}\label{GH4.8}
\Gamma_{GH4}=-\kappa^4\omega^2 L\int d^4x\left\lbrack\frac{1}{64}(\bF^2)^2+\frac{5}{96}\bF^4 \right\rbrack.
\end{equation}

\section{Complete pole part of the effective action}\label{VD6}

The behaviour of the coupling constants in quantum field theory at different energy, or length, scales is governed by the Callan-Symanzik~\cite{Callan,Symanzik}, or renormalization group equations. We will use `t~Hooft's~\cite{tHooftRG} approach as it is based on dimensional regularization.

We can now combine the results for the gauge and ghost fields found above to obtain the complete pole part of the effective action that involves terms only in the background electromagnetic field and deduce the necessary renormalization counterterms. From (\ref{VD4.35}) and (\ref{VD5.16}) we find the quadratic terms to be given by
\begin{equation}\label{6.1}
\Gamma_2=-\frac{\kappa^2\alpha\Lambda}{8\pi^2(n-4)} \int d^4x\bF^2 +\frac{\kappa^2\bar{\beta}}{16\pi^2(n-4)} \int d^4x\left(\partial^\mu\bF_{\mu\nu}\right)^2,
\end{equation}
where $\alpha$ was given in (\ref{VD4.36}) and $\bar{\beta}=\beta+\omega/6$ with $\beta$ given by (\ref{VD4.34}). We have substituted for $L$ from (\ref{A13}).

The quartic pole part of the effective action follows from (\ref{G4.1}) and (\ref{GH4.8}) as
\begin{equation}\label{6.2}
\Gamma_4=-\frac{\kappa^4}{8\pi^2(n-4)} \int d^4x\left\lbrack A_{tot}(\bF^2)^2 +B_{tot}\bF^4\right\rbrack,
\end{equation}
where
\begin{eqnarray}
A_{tot}&=&A-\frac{\omega^2}{64},\label{6.3a}\\
B_{tot}&=&B-\frac{5\omega^2}{96},\label{6.3b}
\end{eqnarray}
with $A$ and $B$ given by (\ref{G4.11}) and (\ref{G4.12}) respectively.

We summarize what would be obtained in various popular choices, along with the gauge condition independent result in Table~\ref{table1}. The final row of this table contains the gauge condition independent result. All of the results, including that which gives rise to the running value of the charge, are seen to be gauge condition dependent when calculated using traditional methods.
\begin{table}[htb]
\caption{\label{table1}This shows the results for $\alpha$ and $\bar{\beta}$ in (\ref{6.1}) and for $A_{tot}$ and $B_{tot}$ in (\ref{6.2}) for popular choices of the parameters. The final row shows the correct gauge condition independent result found with $v=1,\ \omega=1$ and $\xi=0$. For all rows other than the final one we take $v=0$ corresponding to the traditional background-field expression and Feynman rules. Choosing $\omega=0$ is usually called the de~Donder or harmonic gauge. The choice $\kappa^2\xi=1$ is usually called the Feynman gauge. }
\begin{ruledtabular}
\begin{tabular}{c|c|c|c|c}
 &$\alpha$&$\bar{\beta}$&$A_{tot}$&$B_{tot}$\\
 \hline
$\omega=\xi=0,\ v=0$&0&-1/12&1/128&-1/32\\
$\omega=0,\kappa^2\xi=1,\ v=0$&0&1/6&31/384&-35/192\\
$\omega=1,\xi=0,\ v=0$&-3/8&3/4&-5/96&0\\
$\omega=\kappa^2\xi=1,\ v=0$&3/8&1&1/64&-25/192\\
$v=\omega=1,\xi=0$&-3/16&27/16&1/1024&-163/768\\
\end{tabular}

\end{ruledtabular}
\end{table}

The renormalization of the background field and charge were given in (\ref{VD3.24}--\ref{VD3.27}). Using this in the bare Maxwell action (\ref{VD3.2}) gives
\begin{equation}\label{RG1}
S_M=\frac{1}{4}Z_A\int d^4x\bF^2.
\end{equation}
Since this must absorb the pole coming from the quadratic part of $\Gamma_2$ above we find
\begin{equation}\label{RG2}
Z_A=1+\frac{\kappa^2\alpha\Lambda}{2\pi^2(n-4)}
\end{equation}
to one-loop order. The standard `t~Hooft~\cite{tHooftRG} analysis applied to (\ref{VD3.24}), starting from $\ell{de_B}/{d\ell}=0$ results in
\begin{equation}\label{RG3}
E\frac{de}{dE}=\frac{1}{2}(n-4)e+\frac{1}{2}\left( E\frac{d}{dE}\ln Z_A\right)e,
\end{equation}
where we have dropped the subscript `$R$' on the renormalized charge, and used the more conventional energy scale $E$ rather than the length scale $\ell$, with $E=1/\ell$. Because the renormalized charge cannot contain any pole terms the second term of (\ref{RG3}) must be finite
as $n\rightarrow4$, and we can identify the renormalization group $\beta$-function as
\begin{equation}\label{RG4}
\beta_e=\lim_{n\rightarrow4}\frac{1}{2}\left( E\frac{d}{dE}\ln Z_A\right).
\end{equation}

We can write
\begin{equation}\label{RG5}
Z_A=1+\frac{{\mathfrak Z}_1}{(n-4)}++\frac{{\mathfrak Z}_2}{(n-4)^2}\cdots,
\end{equation}
for some coefficients ${\mathfrak Z}_1,{\mathfrak Z}_1,\ldots$ that will in a general theory depend on $e,\kappa,\Lambda$. (In our case we have not obtained a dependence on $e$ because we have not coupled the Maxwell field to charged matter. Our analysis will be general here.)
$\kappa$ and $\lambda$ will satisfy renormalization group equations of their own;
however, the analysis that we have presented is not sufficient to determine this. From the Einstein-Hilbert action (\ref{VD3.3}) we can write
\begin{eqnarray}
\kappa_B&=&\ell^{(n-4)/2}(\kappa+\delta\kappa),\label{RG6}\\
\Lambda_B&=&\Lambda+\delta\Lambda,\label{RG7}
\end{eqnarray}
with the counterterms $\delta\kappa$ and $\delta\Lambda$ expressed as a sum of pole terms in (\ref{RG5}). It can be shown (see \cite{ParkerTomsbook} for example)
\begin{eqnarray}
E\frac{d\kappa}{dE}&=&\frac{1}{2}(n-4)\kappa+\beta_\kappa,\label{RG8}\\
E\frac{d\Lambda}{dE}&=&\beta_\Lambda,\label{RG9}
\end{eqnarray}
for renormalization group functions $\beta_\kappa$ and $\beta_\Lambda$.

To one-loop order, we find from (\ref{RG5}) using (\ref{RG4},\ref{RG8}) and (\ref{RG9})
\begin{eqnarray}
E\frac{d}{dE}\ln Z_A&=&\frac{1}{(n-4)}E\frac{d}{dE}{\mathfrak Z}_1+\cdots\nonumber\\
&=&\frac{1}{(n-4)}\left\lbrace \left(E\frac{de}{dE} \right)\frac{\partial}{\partial e} +\left(E\frac{d\kappa}{dE} \right)\frac{\partial}{\partial\kappa} + \left(E\frac{d\Lambda}{dE} \right)\frac{\partial}{\partial\Lambda} \right\rbrace{\mathfrak Z}_1+\cdots\nonumber\\
&=&\frac{1}{2}e\frac{\partial}{\partial e}{\mathfrak Z}_1+ \frac{1}{2}\kappa\frac{\partial}{\partial\kappa}{\mathfrak Z}_1+\cdots\label{RG10}
\end{eqnarray}
where in the last line we have dropped terms that vanish as $n\rightarrow4$. Comparison of (\ref{RG4}) with (\ref{RG3}) shows that (to one-loop order)
\begin{equation}\label{RG11}
\beta_e=\frac{1}{4}e^2\frac{\partial}{\partial e}{\mathfrak Z}_1+ \frac{1}{4}\kappa e\frac{\partial}{\partial \kappa}{\mathfrak Z}_1.
\end{equation}The first term is that present in the absence of gravity which arises in standard Minkowski spacetime quantum field theory. The second term is a consequence of quantum gravity corrections that we will call $\beta_{grav}$. Using (\ref{RG2}) for ${\mathfrak Z}_1$ we see that
\begin{equation}\label{RG12}
\beta_{grav}=\frac{\alpha}{4\pi^2}\kappa^2 e\Lambda.
\end{equation}
The main calculations presented in this paper show that $\alpha=-3/16$. (See the final line of Table~\ref{table1}.) This means that $\beta_{grav}$ has the opposite sign to the cosmological constant $\Lambda$. We can conclude that if $\Lambda>0$, as current observations favour~\cite{Wmap}, then $e$ is a monotonic decreasing function of $E$. Thus as $E\rightarrow\infty$, meaning that we look at the high energy (short distance) behaviour of the theory, the charge decreases. The quantum gravity correction tends to make the theory asymptotically free. This is also the conclusion found by Robinson and Wilczek~\cite{RobWilczek} for $\Lambda=0$, but the scaling behaviour is very different here. Of course if we use the currently determined values for $\kappa$ and $\Lambda$ then the magnitude of $\beta_{grav}$ is exceptionally small, and the observability of the quantum gravity correction to the running charge is highly unlikely.

For a more realistic gauge theory, if we assume that the quantum gravity correction is the same form as that found in the Maxwell case, then the renormalization group equation for the gauge coupling constant $g$ would be expected to be of the form
\begin{equation}\label{RG13}
E\frac{dg}{dE}=ag^3+bg,
\end{equation}
for calculable expressions $a$ and $b$. $a$ would be the result found in standard Minkowski spacetime calculations, and $b$ would be the correction due to quantum gravity. $b$ would depend on $\kappa^2\Lambda$. Conventionally $b=0$ and asymptotic freedom is determined by the sign of $a$; $a<0$ signals asymptotic freedom~\cite{GrossandWilczek,Politzer} (as in pure Yang-Mills theory, or Yang-Mills theory with not too many fermions) whereas $a>0$ signals the breakdown of a perturbative calculation (as in QED). This raises the intriguing possibility that there could be an ultraviolet fixed point $g=g_\star$ away from zero where
\begin{equation}\label{RG14}
g_\star^2=-b/a.
\end{equation}
This obviously requires $b$ and $a$ to have opposite signs. If the calculation of the present paper applies to matter fields other than Maxwell, it suggests that since $b<0$ a fixed point $g_\star$ will exist if $a>0$. This corresponds to a theory that in the absence of gravity is not asymptotically free (eg. QED), but becomes so once gravity is quantized.

\section{Discussion and conclusions}\label{VD7}

We have shown how the presence of a cosmological constant leads to a non-zero result for the renormalization group $\beta$-function and examined the consequences for the gauge coupling constant. We have also worked out the pole parts of the effective action that involve higher order curvature terms, including those of the Lee-Wick form. By performing the calculations in a sufficiently general way we were able to show conclusively that the traditional background-field result leads to gauge condition dependent results, even though the results are still gauge invariant. One way to ensure that gauge condition independence is maintained is to use the Vilkovisky-DeWitt formalism, as we did.

Notwithstanding our comments concerning the quadratic divergences made in the introduction, it is of interest to examine them more fully within the gauge condition independent formalism, and this is currently under investigation.  We are also looking at the implications for other matter fields (see also \cite{RodSchustnew}) and will report on this elsewhere~\cite{MackayToms}. The extension to higher dimensions with the possible lowering of the energy scale as discussed in \cite{Gogoladze} for the Robinson-Wilczek~\cite{RobWilczek} calculation is of interest. It is also of direct interest to see how the gauge condition dependence cancels in a general gauge by inclusion of the term $T^{k}_{ij}$ in the connection, and this is currently under investigation. 
\begin{acknowledgments}
Some of the more tedious calculations in this paper were done using Cadabra~\cite{cadabra1,cadabra2,cadabra3}. I am very grateful to Kasper Peeters for his help in answering my questions about Cadabra.
\end{acknowledgments}

\appendix*
\section{Evaluation of integrals}

We will give a brief outline of how we may evaluate the products of Green functions encountered in the calculation of the pole part of the effective action described in the main text.  As an example, we will consider
\begin{equation}\label{A1}
I(x,x')=G_{\alpha\beta\gamma\delta}(x,x')\partial_{\mu_1}\cdots\partial_{\mu_r} G_{\lambda\sigma\rho\tau}(x,x')
\end{equation}
where $r=0,1,2$ counts the number of derivatives.  The first step is to use the Fourier expansion (\ref{VD4.13}) to obtain,
\begin{equation}\label{A2}
I(x,x')=\int \frac{d^np}{(2\pi)^n}e^{ip\cdot(x-x')}I(p),
\end{equation}
where
\begin{equation}\label{A3}
I(p)=\int\frac{d^nq}{(2\pi)^n}(iq_{\mu_1})\cdots(iq_{\mu_r}) G_{\lambda\sigma\rho\tau}(q) G_{\alpha\beta\gamma\delta}(p-q).
\end{equation}
We can use the momentum space graviton propagator (\ref{VD4.15}) for each of the two terms in (\ref{A3}).  We will end up with momentum integrals that involve factors of $q_\mu $ in the numerator and various denominators that involve $(q^2-2\lambda),\lbrack(p-q)^2-2\lambda\rbrack$ etc.  At this stage the standard procedure is to introduce Feynman-Schwinger parameters~\cite{Feynman49b} to combine the products of functions in the dominator into a single term, shift the momentum integration accordingly, compute the momentum integration, and finally evaluate the parameter integration.  This process proves to be extremely complicated as the number of factors in the denominator increases when three and four Green functions are present.  Although this, or some equivalent procedure, is necessary for obtaining the finite part of the effective action, a simpler process may be used to obtain the pole terms.  This is because if we are only after the logarithmic divergences of the various integrals over momentum $q$ we only require terms in the integrand that behave like $q^{-4}$ for large $q$.  We may therefore expand the momentum integrands in powers of $q^{-1}$ for large $q$ and extract the term that behaves like $q^{-4}$.  For example, in (\ref{A3}) we use the momentum space expressions for the propagators (\ref{VD4.15}) and expand the product of the two Green functions in powers of $q^{-1}$ keeping the term of order $ q^{-4-r}$ (since there are $r$ factors of $q$ in the numerator).  All of the resulting integrals are then of the form
\begin{equation}\label{A5}
I_{\mu_1\cdots\mu_{2s}}=\int\frac{d^nq}{(2\pi)^n}\frac{q_{\mu_1}\cdots q_{\mu_{2s}}}{(q^2)^{n/2+s}},
\end{equation}
where $s=0,1,2,\ldots$.  When the number of factors of $q_\mu$ in the numerator is odd we regularize the result to zero since the integrand is an odd function of $q$.

$I_{\mu_1\cdots\mu_{2s}}$ is a symmetric tensor, and we can write
\begin{equation}\label{A6}
I_{\mu_1\cdots\mu_{2s}}=f(n,s)\delta_{\mu_1\cdots\mu_{2s}}
\end{equation}
for some function $f(n,s) $ with $\delta_{\mu_1\cdots\mu_{2s}}$ expressible as the sum of products of $s$ Kronecker deltas with all possible pairings of indices.  For example,
\begin{equation}\label{A7}
\delta_{\mu_1\mu_2\mu_3\mu_4}=\delta_{\mu_1\mu_2}\delta_{\mu_3\mu_4} +\delta_{\mu_1\mu_3}\delta_{\mu_2\mu_4} +\delta_{\mu_1\mu_4}\delta_{\mu_2\mu_3}.
\end{equation}
it is easy to see that
\begin{equation}\label{A8}
\delta^{\mu_1\mu_2}\delta_{\mu_1\cdots\mu_{2s}}=(n+2s-2)\delta_{\mu_3\cdots\mu_{2s}}.
\end{equation}
From (\ref{A5}), it is clear that
\begin{equation}\label{A9}
\delta^{\mu_1\mu_2}I_{\mu_1\cdots\mu_{2s}}=I_{\mu_3\cdots\mu_{2s}}.
\end{equation}
Using (\ref{A6}) and (\ref{A8}) shows that
\begin{equation}\label{A10}
(n+2s-2)f(n,s)=f(n,s-1).
\end{equation}
This allows us to relate all integrals of the form (\ref{A5}) to the basic logarithmically divergent integral
\begin{equation}\label{A11}
L=\int\frac{d^nq}{(2\pi)^n}\frac{1}{(q^2)^{n/2}}.
\end{equation}
If we are interested in the case, $n\rightarrow4$, then,
\begin{equation}\label{A12}
f(4,s)=\frac{L}{2^s(s+1)!}
\end{equation}
and
\begin{equation}\label{A13}
L=-\frac{1}{8\pi^2(n-4)}.
\end{equation}
Other spacetime dimensions are easily dealt with.  For the quadratic part of the effective action we have checked that this procedure agrees with the method of combining denominators using Feynman-Schwinger parameters~\cite{Feynman49b} and found the procedure just described much easier to implement.

Proceeding as described will yield a result for $I(p)$ that has the basic logarithmic pole in $L$ with various factors of Kronecker deltas and momenta $p_\mu$. When used back in expressions like (\ref{A2}) the result is expressible as Dirac delta functions and derivatives of Dirac delta functions. Integration over the spacetime coordinates in the effective action removes the Dirac deltas and their derivatives and places the derivatives on the background gauge field. This is how all of the pole terms obtained in the present paper were evaluated. Although the calculations are tedious they are straightforward, and we omit all such technical details for brevity. Many of the calculations were done with or checked with Cadabra~\cite{cadabra1,cadabra2,cadabra3}.

\end{document}